%% file: main.tex
\documentclass[letterpaper,times]{IONconf}

\usepackage[dvipsnames]{xcolor}
\usepackage{soul}
\colorlet{r1}{ForestGreen!0}
\colorlet{r2}{yellow!0}
\colorlet{r3}{RoyalPurple!20}
\newcommand{\Hl}[2][\empty]{%
\ifx#1\empty
\else
\sethlcolor{#1}%
\fi
\hl{#2}}
\usepackage{soul,color}
\soulregister\Hl{7}
\soulregister\ref7
\soulregister\eqref7
\soulregister\textit7
\soulregister\defemph7
\soulregister\textrm7
\soulregister\cite7
\soulregister\citep7

\usepackage{hyperref} 
\usepackage{amsmath}
\usepackage[]{graphicx}
\usepackage{float}
\usepackage{subcaption}
\usepackage{mathtools}
\usepackage{amssymb}

\usepackage{upgreek}

\usepackage{bbm}

\usepackage[backend=biber,style=apa, natbib=true]{biblatex}
\usepackage{verbatim}
\usepackage{bm}
\usepackage{multirow}

\usepackage[ruled,vlined,linesnumbered]{algorithm2e}

\SetCommentSty{mycommfont}
\SetKwInput{KwInput}{Initialize} 
\SetKwInput{KwOutput}{Output at any $t$th time iteration} 

\usepackage{siunitx}

\usepackage{outlines}
\usepackage{enumitem}
\setenumerate[2]{label=\alph*.}
\setenumerate[3]{label=\roman*.}

\include{macros}

\title{\huge Designing Low-Correlation GPS Spreading Codes with a Natural Evolution Strategy Machine Learning Algorithm}
\author{Tara~Yasmin~Mina  and Grace~Xingxin~Gao, \textit{Stanford~University}}

\makeatletter 
\newcount\SOUL@minus
\makeatother  
\date{}

\begin{document}
\maketitle
\pagestyle{plain}

\input{sections/abstract}

\input{sections/introduction}
\input{sections/preliminaries}
\input{sections/method}
\input{sections/experiments}
\input{sections/conclusion}
\input{sections/appendix}

\input{sections/acknowledgements}

\printbibliography[heading=bibintoc, title={References}]


\end{document}

%% file: macros.tex
\renewcommand{\normalsize}{\fontsize{11}{12}\selectfont}

\newcommand{\defemph}[1]{\emph{#1}}



%% file: sections/abstract.tex
\section*{Abstract}

With the birth of the next-generation GPS III constellation and the upcoming launch of the Navigation Technology Satellite-3 (NTS-3) testing platform to explore future technologies for GPS, we are indeed entering a new era of satellite navigation. 
Correspondingly, it is time to revisit the design methods of the GPS spreading code families. 
In this work, we develop a natural evolution strategy (NES) machine learning algorithm with a Gaussian proposal distribution which constructs high-quality families of spreading code sequences. 
{We minimize the maximum between the mean-squared auto-correlation and the mean-squared cross-correlation and demonstrate the ability of our algorithm to achieve better performance than well-chosen families of equal-length Gold codes and Weil codes, for sequences of up to length-1023 and length-1031 bits and family sizes of up to 31 codes.
Furthermore, we compare our algorithm with an analogous genetic algorithm implementation assigned the same code evaluation metric. 
To the best of the authors' knowledge, this is the first work to explore using a machine learning approach for designing navigation spreading code sequences.}

%% file: sections/introduction.tex
\section{Introduction}\label{sec:intro}

On January 13th of 2020, the U.S. Air Force 2nd Space Operations Squadron (2 SOPS) issued a statement that the first GPS~III satellite was marked healthy and available for use~\citep{announcementGPSIII}. This announcement officially marked the birth of the next-generation GPS constellation. In addition to broadcasting the new L1C signal, the modernized constellation is distinguished by its reprogrammable payload, which allows it to evolve with new technologies and changing mission needs. 

\begin{figure}
	\centering
	\includegraphics[width=0.7\textwidth]{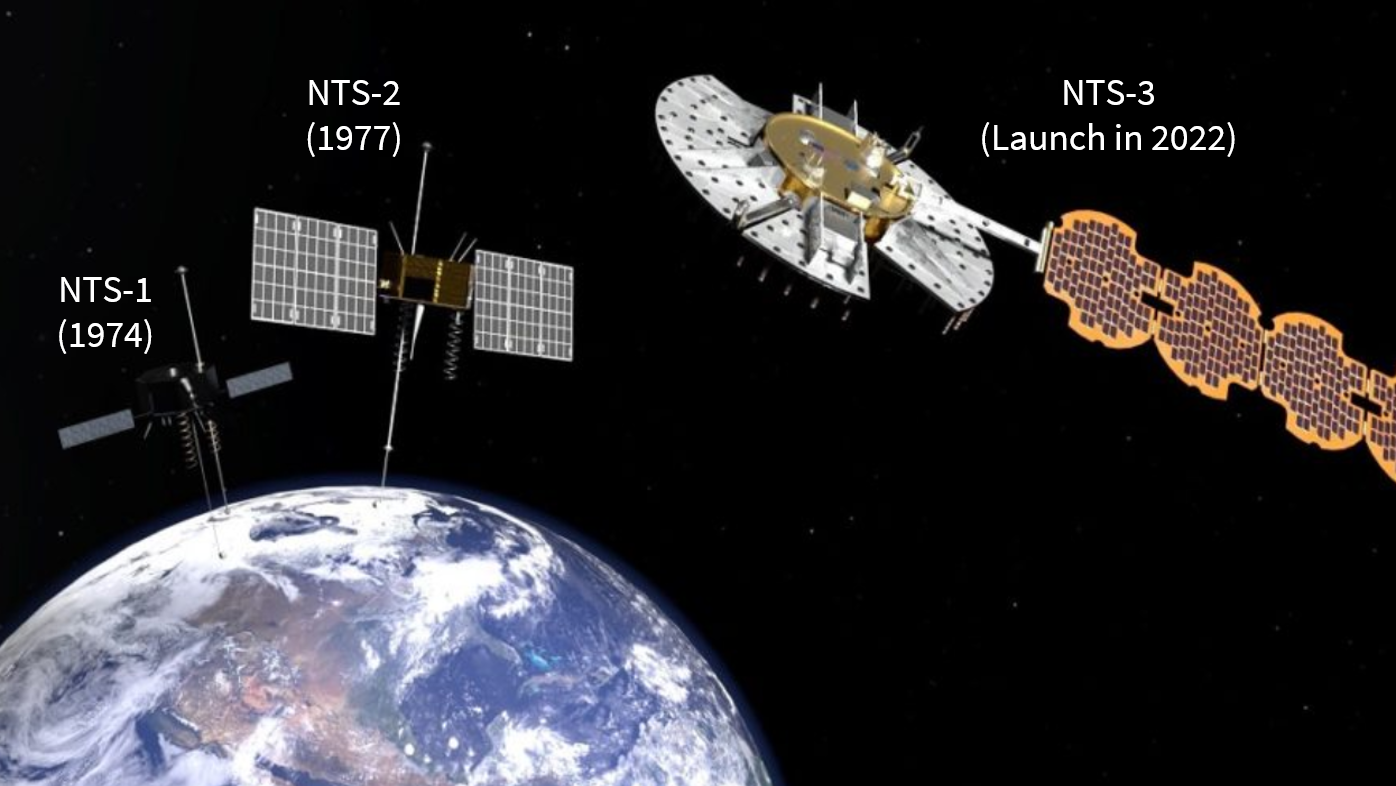}
	\caption{Illustrations of the satellite testing platforms used in each of three Navigation Technology Satellite (NTS) initiatives. Illustration credit: Lt. Jacob Lutz, AFRL Space Vehicles Directorate.}
	\label{fig:nts}
\end{figure}

Furthermore, with the upcoming launch of the Navigation Technology Satellite-3 (NTS-3) testing platform in {2023~\citep{nts3delay}}, the United States Air Force (USAF) seeks to explore technologies which will help shape future GPS constellations~{\citep{nts3insideGNSS}}. NTS-3 will demonstrate the agility of the next-generation satellite-based navigation architecture and the ability to rapidly deploy new technological advancements and capabilities via the reprogrammable nature of the upcoming GPS system. Indeed, this is the third Navigation Technology Satellite (NTS) mission, with the previous two, NTS-1 and NTS-2, developed in the 1970s in order to validate technologies, including the rubidium and cesium atomic clocks~\citep{ntsGPSWorld}, that were integrated into the first generation of GPS satellites. Illustrations of the three NTS testing platforms are shown in Fig.~\ref{fig:nts}. 

The NTS-3 program will further test several new technologies, including new signal designs for improved GPS security and interference mitigation~\citep{nts3insideGNSS}. According to a Request for Information announcement~{\citep{afrlRFI,RFIInside2014}}, the Air Force Research Lab (AFRL) has expressed interest in exploring modifications to all layers of the GPS signal in order to enhance PNT resiliency and performance. As we enter a new era of satellite navigation, we believe it is time to revisit the design methods of the GPS spreading code families. Correspondingly, in this work, we leverage recent advances in machine learning techniques to explore a new platform for learning high-quality spreading codes via a natural evolution strategy (NES) machine learning framework.

\subsection{Advantages of Using Memory Codes for GPS}
Currently, all broadcast GPS signals use algorithmically generable spreading code sequences. The legacy L1 C/A signal uses Gold codes~\citep{gold1967optimal}, while the modernized L1C signal, to be broadcast on the GPS~III satellites, uses Weil codes~\citep{guohua2002pseudonoise,rushanan2006weil}, along with a 7-bit extension in order to satisfy the desired spreading code length of 10,230 bits. Having algorithmically generable sequences was an important design consideration during the initial development phase of GPS in the 1970s due to memory storage constraints, as well as computation limitations which restricted the ability to search for better sequences. However, memory has become an increasingly inexpensive resource, allowing receivers to store complete families of spreading code sequences~--~though it should be acknowledged that currently, memory is still one of the main influencers of receiver chip size~\citep{vanDiggelenInside2014,van2020high}. Furthermore, over the past several decades, processing power has also significantly increased, which correspondingly has allowed the field of machine learning and related research advancements to flourish~\citep{lu2017ai}. 

Lower bounds on the maximum periodic non-central auto- and cross-correlation value for spreading code families have been analytically derived in prior work~\citep{stular2000maximum}, including the Welch bound~\citep{welch1974lower}, the Sarwate bound~\citep{sarwate1979bounds}, and the Sidelnikov bound~{\citep{sidelnikov1971cross,sidel1971mutual}}. The Welch bound is valid for any sequence with symbols of a fixed norm, including potentially complex spreading codes, while the Sidelnikov bound is a tighter lower bound for sequences with symbols that are roots-of-unity. As a result, for binary spreading codes, which fall into this roots-of-unity classification with symbols in the set~$\{+1, -1\}$, the Sidelnikov bound provides a tighter lower bound than the Welch bound on the maximum periodic correlation performance. 

Gold code families are known to be asymptotically optimal with respect to the Sidelnikov bound, as the code length increases~\citep{burroughs1996performance}. However, this optimality characteristic only applies for the \textit{specific sequence length} and the \textit{complete set of code sequences} provided by the Gold code family definition. For the length-1023 bit Gold codes, for example, this optimality characteristic only applies for the complete Gold family size, made up of over 1000 spreading codes in total~\citep{gold1967optimal}. Indeed, the lower bounds on the maximum correlation sidepeak increase with the size of the spreading code family~{\citep{welch1974lower,sarwate1979bounds,sidelnikov1971cross,sidel1971mutual}}, suggesting that for the same sequence length, better performance could be found for a smaller code family size. Furthermore, there is debate as to whether the maximum correlation sidepeak is an ideal metric for spread spectrum system applications~{\citep{pursley1977performance,stular2001mean,ganapathy2011new2,ganapathy2011new}}. Indeed, the multiple access interference (MAI) of a spread spectrum system, i.e. the signal noise caused by inter-signal interference within the shared channel, relates to the \textit{average} correlation performance rather than the worst-case.

Spreading codes which are not algorithmically generable are commonly called \textit{memory codes}, since these codes must be stored in memory. Memory codes are not limited to specific lengths of code sequences or specific code family sizes. Indeed, designing a nonconforming spreading code length, as was done for the GPS L2C, L5, and L1C signals, would require truncation or extension of the algorithmically generable sequences, which disturb the coveted correlation properties of these codes~{\citep{wallner2007galileo}}. As a result, by expanding the design space to the set of all binary sequences, we have a greater range of possible code families and a better opportunity to find a superior set of codes; although, the exponentially larger design space also greatly complicates the code design method.

\subsection{Related Prior Work}
Several past works have designed memory codes using genetic algorithms~(GAs), including for the E1 Open Service~(OS) and the E6 Commercial Service~(CS) signals of the Galileo satellite constellation~\citep{wallner2007galileo,interface2021galileo}. Genetic algorithms~\citep{holland1975adaptation} are optimization algorithms which mimic the process of biological evolution. In particular, GAs maintain a population of design points, while selecting and recombining high-performing candidate solutions at each iteration. 

Genetic algorithms were utilized to design spreading codes for Galileo which exhibit the \textit{Autocorrelation Sidelobe Zero} property, where the auto-correlation is 0 at a relative delay of $\pm 1$~chip~\citep{wallner2007galileo}. GAs were also utilized to design spreading code families for improved indoor positioning applications~{\citep{avila2006optimize}}. Work has also been done to define several cost parameters for selecting high quality GNSS spreading codes~{\citep{soualle2005spreading,winkel2011spreading}}, including evaluating the mean squared auto- and cross-correlation components above the Welch bound~\citep{welch1974lower}. Additionally, in our prior work, we designed a multi-objective GA platform for designing navigation spreading codes which improved on two objectives simultaneously: the mean absolute auto-correlation and cross-correlation of the code family~\citep{mina2019devising}. Learning-based techniques have also been explored to design Error Correcting Codes~(ECCs)~{\citep{huang2019ai}}, which are binary sequences that encode messages to improve the detection of communication transmission errors. In their work, Huang et al. explore policy gradient and advantage actor critic methods to design ECCs by optimizing the code set performance with regards to the block error rate~\citep{huang2019ai}.

\subsection{Objective and Key Contributions}

In this work, we seek to explore using a machine learning technique for the application of navigation spreading code design. This work is based on our recent ION GNSS+ 2020 conference paper~\citep{mina2020designing}, and to the best of our knowledge, this is the first work which explores using a machine learning approach to design navigation spreading codes. In particular, the key contributions of our work are the following:
\begin{enumerate}
	\item We develop a \textbf{natural evolution strategy (NES)} machine learning algorithm which constructs high-quality families of spreading code sequences.
	\item We utilize a \textbf{Gaussian proposal distribution} parameterized by a \textbf{generative neural network (GNN)} in order to model a search distribution over the design space of possible binary spreading code families.
	\item We incorporate a baseline to reduce the variance of the NES gradient estimate and to improve the rate of learning.
	\item We use a \textbf{maximization evaluation metric} to ensure the algorithm minimizes both the auto-correlation and cross-correlation characteristics of the spreading codes simultaneously.
\end{enumerate}

With our algorithm, we demonstrate the ability to achieve low auto- and cross-correlation side peaks within the family of spreading codes. We further compare the correlation performance of the learned spreading codes with those of well-chosen families of equal-length Gold codes and Weil codes as well as with an analogous genetic algorithm implementation assigned the same code evaluation metric as our proposed algorithm.

\subsection{Paper organization}
The remainder of the paper is organized as follows: Section~\ref{sec:tech_back} provides relevant technical background for the paper; Section~\ref{sec:algo} describes our NES machine learning algorithm for the design of spreading code families; Section~\ref{sec:exp_val} presents our experimental validation setup and results; Section~\ref{sec:conc} concludes this paper; and Section~\ref{sec:app} provides additional background on artificial neural networks.

%% file: sections/preliminaries.tex
\section{Background}\label{sec:tech_back}
\subsection{Binary Sequence Representation}
For mathematical convenience, we represent a binary sequence~$\bar{x}^{(0,1)}$ with elements of the set~$\{0,1\}$ as a sequence~$\bar{x}$ with elements of the set~$\{+1, -1\}$ by using the following conversion
\begin{align}
	\forall i &: \bar{x}^{(0,1)}(i) = 0, \bar{x}(i) = +1 \nonumber \\ 
	\forall i &: \bar{x}^{(0,1)}(i) = 1, \bar{x}(i) = -1 \ , \label{eq:bin_seq_rep}
\end{align}
where~$\bar{x}^{(0,1)}(i)$~and~$\bar{x}(i)$ represent the $i^{th}$ element of the two binary sequence representations. In this paper, unless otherwise specified, we represent sequences using the $(+1, -1)$ binary representation. To denote sequences with the $(0,1)$ binary representation, we utilize an additional superscript of $(0,1)$, i.e.~$\bar{x}^{(0,1)}$.

\begin{figure}
	\centering
	\includegraphics[width=0.8\textwidth]{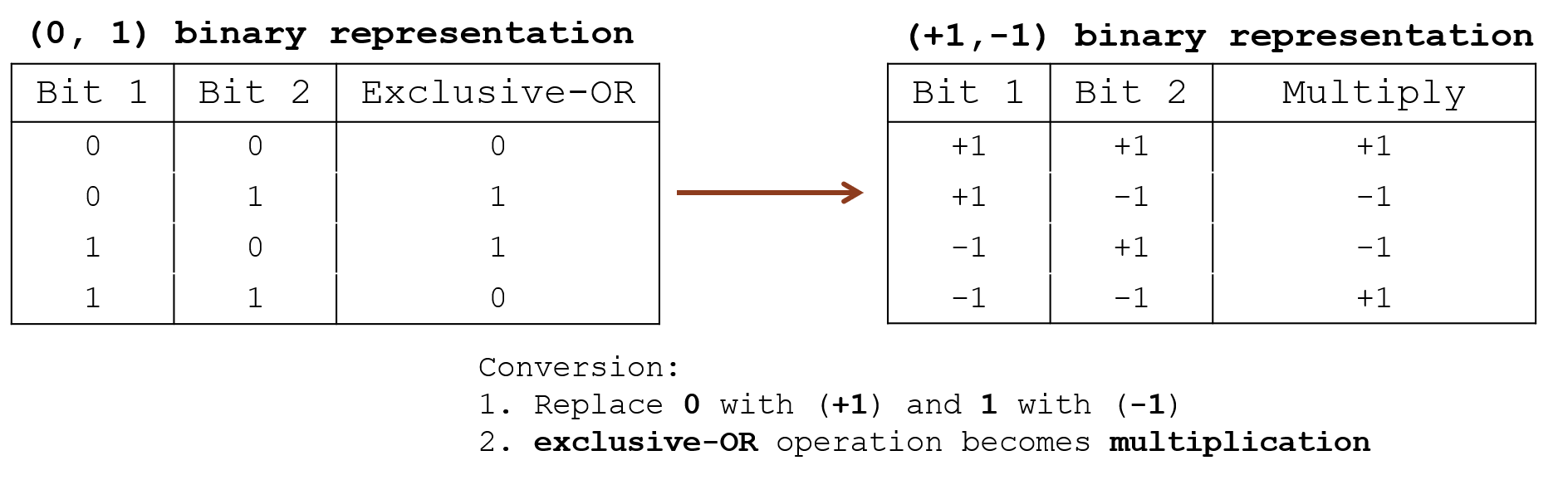}
	\caption{Illustration of the conversion from the $(0,1)$~binary sequence representation to the $(+1, -1)$~representation used for mathematical convenience. Consequently, the exclusive-OR operation between elements of the $(0,1)$~representation becomes a multiplication operation.}
	\label{fig:xor_convert_mult}
\end{figure}

The $(+1,-1)$~representation of the binary sequences simplifies the auto-correlation and cross-correlation computations performed on one or more binary sequences. In particular, an exclusive-OR operation, i.e.~$b_1 \oplus b_2$, between two elements of the $(0,1)$~binary representation becomes a simple multiplication operation, i.e.~$b_1 \cdot b_2$, between two elements of the $(+1, -1)$~binary representation, as illustrated in Fig.~\ref{fig:xor_convert_mult}.

\subsection{Periodic Auto- and Cross-Correlation}\label{ssec:corr_def}
Using the $(+1, -1)$~binary sequence representation defined in Eq.~\eqref{eq:bin_seq_rep}, let $\overline{x}_k$ represent the $k^{th}$~binary sequence in a family of length-$\ell$ spreading code sequences. Let $\overline{x}_k(i) \in \{+1, -1\}$ represent the $i^{th}$~element of the sequence, where~$i \in \mathbb{Z}\cap[0, \ell-1]$ and $\mathbb{Z}$ represents the set of all integer values. Because $\overline{x}_k$ is a periodic sequence, we have that~$\overline{x}_k(i)=\overline{x}_k(i+\ell)$. The normalized, {periodic even auto-correlation} of sequence~$k$ at a relative delay of~$\delta$~bits is defined as 
\begin{align}
	R_k(\delta) &\coloneqq \frac{1}{\ell} \sum_{i=0}^{\ell-1} \overline{x}_k(i) \overline{x}_k(i-\delta) \label{eq:auto_corr_def} \ .
\end{align}
Similarly, in Eq.~\eqref{eq:cross_corr_def} we define the normalized, periodic even cross-correlation between sequences~$k$~and~$m$ at a relative delay of $\delta$~bits as
\begin{align}
R_{k,m}(\delta) &\coloneqq \frac{1}{\ell} \sum_{i=0}^{\ell-1}  \overline{x}_k(i) \overline{x}_m(i-\delta) \ . \label{eq:cross_corr_def}
\end{align}
The even correlation is distinguished from the \textit{odd correlation}, which defines the correlation in the case that a data flip occurs within the integration period of the correlators~{\citep{winkel2011spreading,prnFamilyinsideGNSS}}. Correspondingly, the odd correlation for sequence~$k$ would be computed by flipping the sign of the second multiplicative terms of Eq.~\eqref{eq:auto_corr_def}~and~\eqref{eq:cross_corr_def}, i.e.~$\overline{x}_k(i - \delta)$ and $\overline{x}_m(i - \delta)$ respectively, when~$i < \delta$.

\subsection{Algorithmically Generable GPS Spreading Codes}
\begin{figure}
	\centering
	\includegraphics[width=0.6\textwidth]{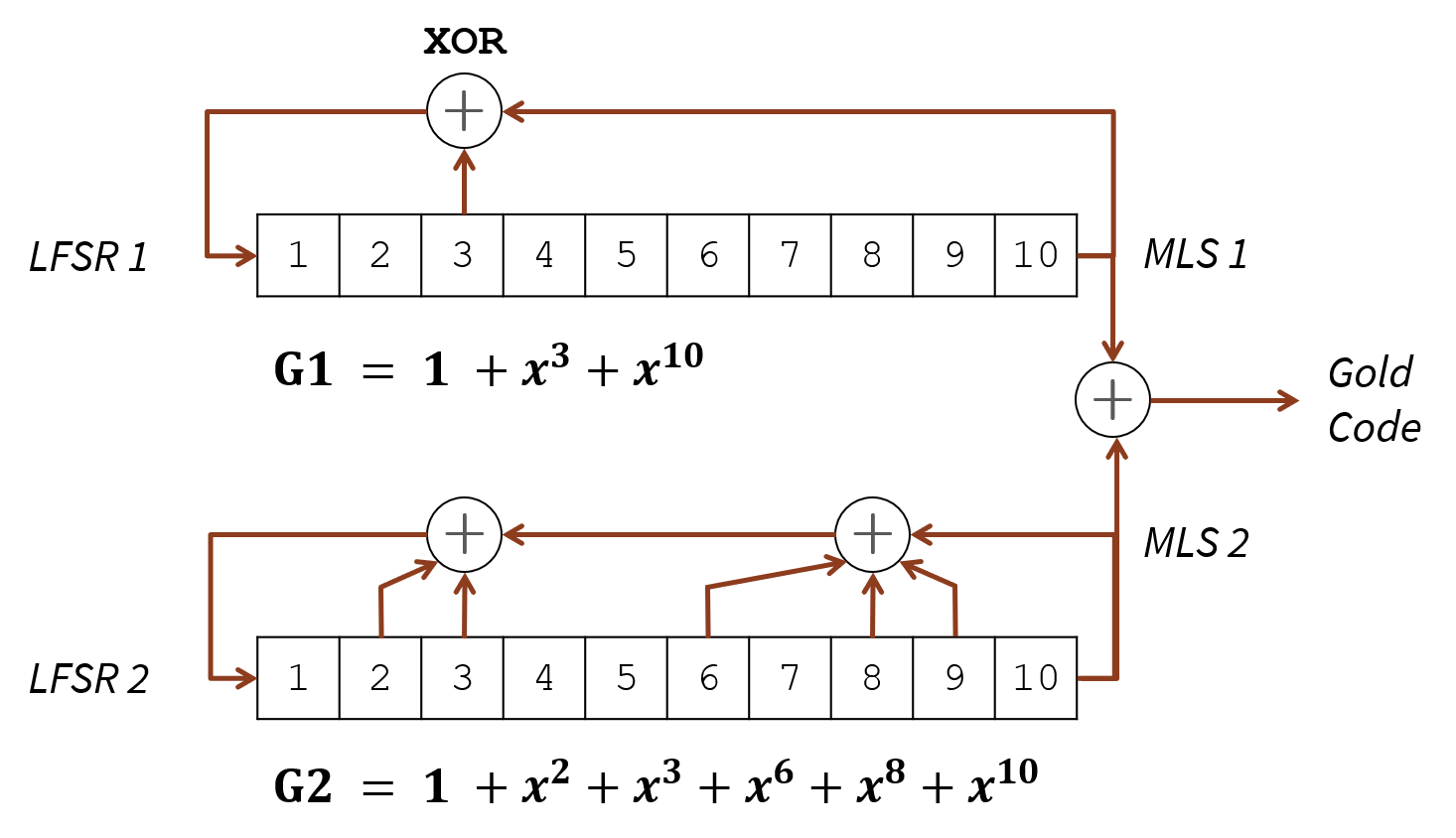}
	\caption{Diagram of the linear feedback shift register (LFSR) pair with their corresponding characteristic equations ($G1$ and $G2$) utilized to generate length-1023 bit Gold codes used in GPS L1 C/A~\protect\citep{interface2021gps}. Only two 10-bit LFSRs are required for the generation of the GPS L1 C/A spreading codes.}
	\label{fig:goldCodes}
\end{figure}

Currently, all broadcast GPS signals use algorithmically generable spreading code sequences. The legacy L1 C/A GPS signal, for example, uses length-1023 Gold codes, which can be generated with two 10-bit linear feedback shift registers~(LFSRs), as shown in Fig.~\ref{fig:goldCodes}. Gold codes~\citep{gold1967optimal} are a class of binary spreading code families which can be generated from two maximum-length sequences, or $m$-sequences. Maximum-length sequences are pseudorandom sequences of length~$\ell = 2^{n} - 1$, which are generated using an~$n$-bit~LFSR~\citep{golomb1967shift}.
Recently developed for the modernized GPS L1C signal, Weil codes are families of spreading codes which were initially proposed by~\citep{guohua2002pseudonoise} and further analyzed and proposed for the L1C signal by~\citep{rushanan2006weil,rushanan2007spreading}. A Weil code family exists for any prime number sequence length-$p$ and is generated from the corresponding length-$p$ \textit{Legendre sequence}, constructed from the \textit{Legendre symbol}~\citep{legendre1808essai,yan2002number}. In this work, we construct both Gold codes and Weil codes in order to compare the performance of these spreading code families utilized in GPS with the performance of the spreading codes generated from our proposed NES machine learning algorithm.

\subsection{Natural Evolution Strategy Algorithms} \label{ssec:NES}
Natural Evolution Strategies (NES)~{\citep{rechenberg1973evolutionsstrategie,wierstra2008natural,wierstra2014natural,kochenderfer2019algorithms}} are a subset of stochastic optimization methods which model a probability distribution, or a \textit{proposal distribution}, over the optimization design space. Each sample from this proposal distribution outputs a candidate design point. In the context of designing navigation spreading codes, for example, each sample of the proposal distribution would output a set of binary sequences. NES requires choosing a family of parameterized proposal distributions~$\mathcal{P}_\Theta$, explicitly defined as
\begin{align*}
	\mathcal{P}_\Theta = \{ p_\theta(x) : x \in \mathcal{X}, \theta \in \Theta \}
\end{align*}
where~$\Theta$ is the set of possible vector-valued parameters~$\theta$ defining each distribution in the family and where~$\mathcal{X}$ represents the design space over which we are optimizing.
	  
Within the specified family of probability distributions, NES methods directly optimize the proposal distribution~$p_\theta$ with respect to its parameterization~$\theta$, in order to guide the search process towards higher performing regions of the design space. 
NES algorithms are especially useful for non-convex optimization problems with large design spaces, including combinatorial design spaces, which applies to our application of interest with the design of navigation spreading codes. For high-dimensional design spaces, often diagonal covariance matrices are considered for the proposal distribution in order to improve \textit{computational efficiency} with regards to the number of NES parameters to optimize as well as 
\textit{sample efficiency} with regards to the number of required Monte Carlo samples to effectively estimate the search gradient during the optimization step~{\citep{wierstra2014natural,schaul2011high}}. Further variations of traditional NES methods have also been explored to handle highly multi-modal optimization landscapes by leveraging heavy-tailed distributions, including those with undefined variances, such as the Cauchy distribution~\citep{schaul2011high}.

NES methods optimize the proposal distribution within the defined family~$\mathcal{P}_\Theta$, by seeking to minimize the expected objective function~$J$, defined as
\begin{align}
J(\theta) &\coloneqq \mathbb{E}_{x \sim p_\theta}\left[ f(x) \right] \nonumber \\
&= \int p_\theta(x) f(x) dx \label{eq:def_pg_obj} \ ,
\end{align}
where~$x$ represents a particular point in the design space~$\mathcal{X}$, $p_\theta(x)$ represents the probability of selecting $x$ given the parameter~$\theta$, and $f(x)$ represents the objective function value of the design point~$x$, which we seek to minimize. Thus, NES methods seek to find the best proposal distribution by solving the optimization problem
\begin{align}
	\theta^* &= \arg \min_\theta J(\theta) \ . \label{pg_opt_prob}
\end{align}
To optimize the proposal distribution, the gradient of the expected objective function, i.e.~$\nabla_\theta J(\theta)$, can be estimated via Monte Carlo sampling of the objective function using the current proposal distribution~{\citep{glynn1990likelihood,kochenderfer2019algorithms,mohamed2020monte}}:
\begin{align}
	\nabla_\theta J(\theta) &= \nabla_\theta \mathbb{E}_{x \sim p_\theta}\left[ f(x) \right] \nonumber \\
	&= \int f(x) \nabla_\theta p_\theta(x) dx \nonumber \\
	&=   \int f(x) \frac{\nabla_\theta p_\theta(x)}{p_\theta(x)}  p_\theta(x) dx \label{eq:log_deriv_trick_1} \\
	&= \int \left[f(x) \nabla_\theta \log p_\theta(x) \right] p_\theta(x) dx \label{eq:log_deriv_trick_2} \\
	&= \mathbb{E}_{x \sim p_\theta}\left[ f(x) \nabla_\theta \log p_\theta(x) \right] \\
	&\approx \frac{1}{N} \sum_{i=1}^N f(x^i) \nabla_\theta \log p_\theta(x^i) \label{eq:grad_est}
\end{align}
where $N$ represents the number of samples from the proposal distribution and $x^i$ represents the $i^{th}$~sample. Eqs.~\eqref{eq:log_deriv_trick_1}~and~\eqref{eq:log_deriv_trick_2} demonstrate the so-called ``log-derivative trick,'' which is observed in a variety of machine learning contexts, including reinforcement learning and variational inference~\citep{mohamed2020monte}.
\begin{figure}
	\centering
	\includegraphics[width=\textwidth]{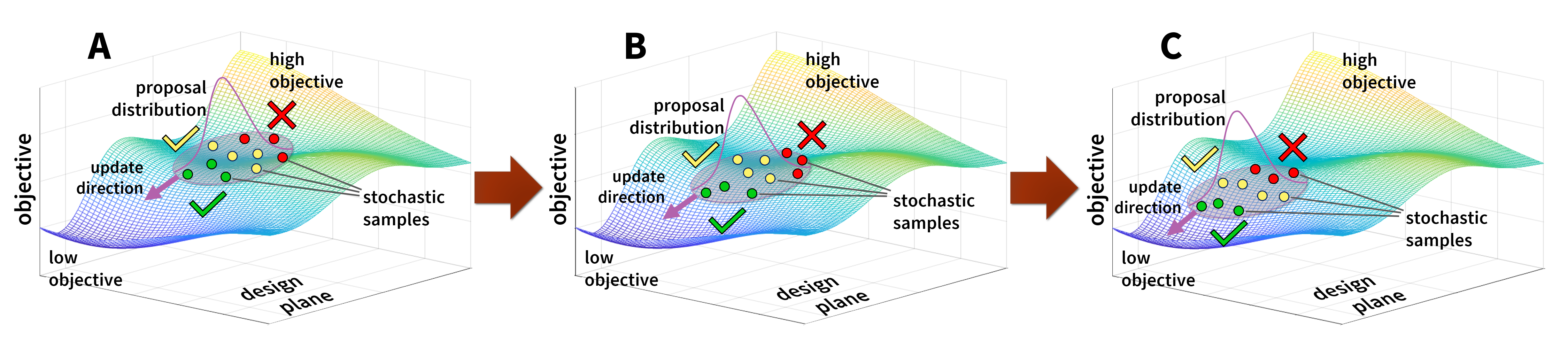}
	\caption{Illustration depicting how the NES algorithm optimizes the expected objective~$J(\theta)$. From instance \textbf{A} to instance \textbf{B}, the NES optimizer updates the parameters of the proposal distribution in order to increase the likelihood of design points which lead to lower values of the objective function (indicated by the green sampled points), while decreasing the likelihood of design points which have higher objective function values (indicated by the red sampled points). This update then guides the proposal distribution down the objective function landscape in instance \textbf{B}, toward regions where the objective is lower. The optimizer again repeats this process from instance \textbf{B} to instance \textbf{C}, thereby further guiding the proposal distribution down the objective function landscape. Indeed, the stochasticity of the proposal distribution induces the NES algorithm to explore the design space and the optimization step iteratively steers the search process towards better performing regions of the design space, where the objective function value is lower. }
	\label{fig:traj_illus}
\end{figure}

From this gradient estimate, using any first-order optimizer (e.g. stochastic gradient descent, Adam~\citep{kingma2014adam}), NES improves its proposal distribution vector-valued parameter~$\theta$ in order to minimize the expected objective function in Eq.~\eqref{eq:def_pg_obj}
\begin{align}
	\theta &\leftarrow \theta - \delta \theta \nonumber \\
	\delta \theta &\coloneqq g_{opt}\left(\nabla_\theta J(\theta)\right)\ , \label{eq:opt_step}
\end{align}
where $g_{opt}$ is the generic first-order optimizer utilized to iteratively improve the proposal distribution. As illustrated in Fig.~\ref{fig:traj_illus}, this optimization process increases the probability of design points which lead to lower values of the objective function~$f$ to be minimized and decreases the probability of those which result in higher values of the objective function. Indeed, the stochasticity of the proposal distribution induces the NES algorithm to explore the high-dimensional design space and further improve its proposal distribution to output better performing design points.

%% file: sections/method.tex
\section{Proposed Natural Evolution Strategy Algorithm for Spreading Code Design} \label{sec:algo}
\subsection{High-Level Learning Framework}
\begin{figure}
	\centering
	\includegraphics[width=0.7\textwidth]{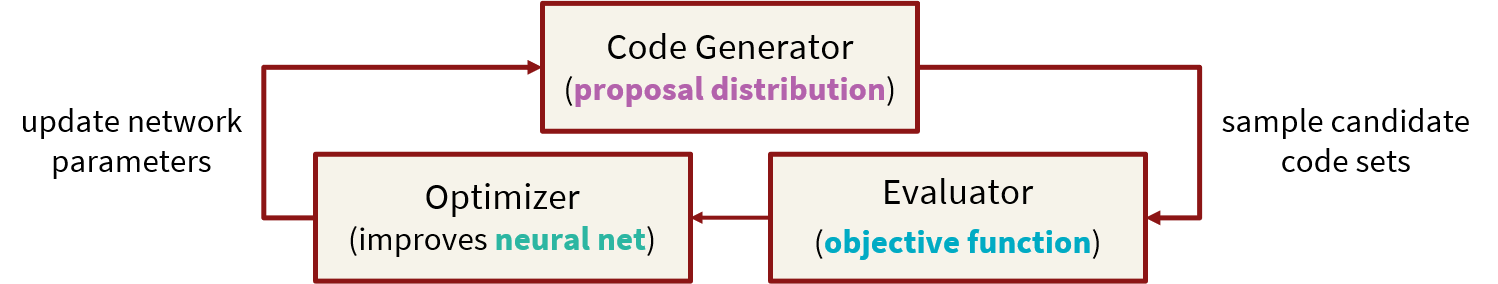}
	\caption{High-level learning framework utilized for spreading code generation. The \textbf{code generator} outputs a set of sampled candidate code sets which are assessed by the \textbf{evaluator}. From the sampled design points and corresponding objective function values, \textbf{optimizer} improves the proposal distribution by updating its parameters as represented by a generative nerual network.}
	\label{fig:rl_framework}
\end{figure}

For the spreading code design application, the machine learning framework utilized is represented at a high-level in Fig.~\ref{fig:rl_framework}. A \textbf{code generator} maintains a proposal distribution over the spreading code design space. Then, an \textbf{evaluator} assesses the performance of the code generator with respect to the objective function~$f$ by sampling from its proposal distribution. The objective function represents an evaluation metric for the spreading code family, based on the auto- and cross-correlation properties of the code set. Finally, as in Eq.~\eqref{eq:grad_est}, the \textbf{optimizer} estimates the gradient of the expected objective function with respect to the parameters of the proposal distribution, as represented by a generative neural network (GNN). The optimizer then updates the network parameters in order to improve the proposal distribution for the code generator.

\subsection{Gaussian Proposal Distribution Representation}
\begin{figure}
	\centering
	\includegraphics[width=0.8\textwidth]{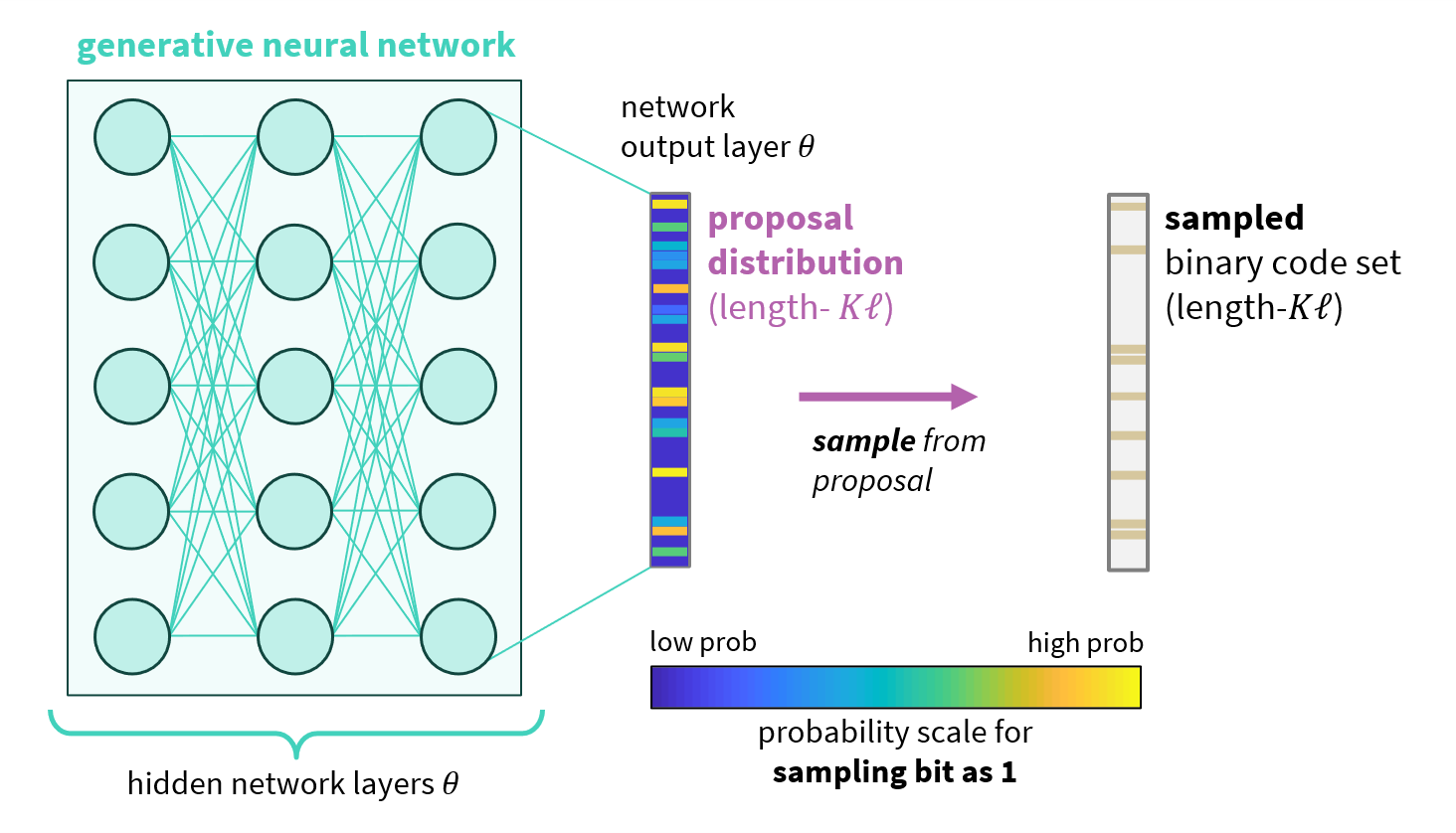}
	\caption{Illustration of \textbf{generative neural network (GNN)} and proposal distribution representation. The GNN outputs the mean vector of the multivariate Gaussian \textbf{proposal distribution}, from which a sampled output code set for the complete spreading code family is generated. In particular, the sampled output code set is a binary vector of length-$K\ell$ bits, where~$K$ denotes the number of codes in the family and~$\ell$ denotes the length of each spreading code sequence. With this representation of the proposal distribution, the network parameters correspond to the NES parameters~$\theta$, which are optimized during training.}
	\label{fig:policy_network_rep}
\end{figure}

Using a neural network architecture, we represent a probability distribution over the design space of all possible binary spreading code families. Thus, the parameters of the proposal distribution~$\theta$ correspond to the hidden layers of the network. As depicted in Fig.~\ref{fig:policy_network_rep}, for the code generator to output a set of binary codes, it must sample from this \textit{proposal distribution} provided by the GNN, with the currently learned network parameters~$\theta$. For a concise introduction to artificial neural networks, we would suggest the reader to reference Sec.~\ref{sec:app}.

\begin{figure}
	\centering
	\includegraphics[width=0.55\textwidth]{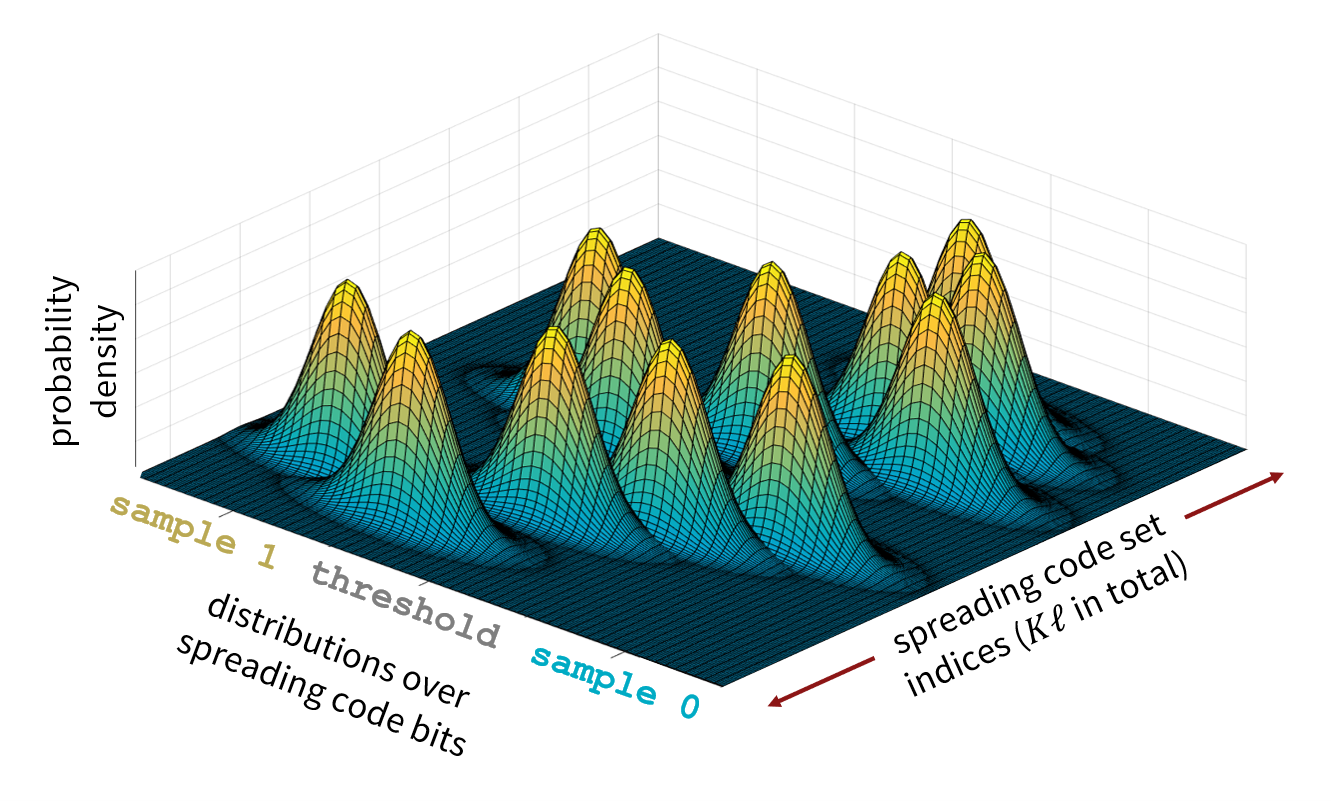}
	\caption{Illustration of the proposal distribution utilized for spreading code design. We model the code design proposal distribution as an uncorrelated multivariate Gaussian family, where each component corresponds to an index in the family of binary spreading codes. Correspondingly, the total length of the spreading code set is~$K\ell$, where~$K$ denotes the number of codes in the family and $\ell$ denotes the length of each binary spreading code sequence.}
	\label{fig:gauss_pg}
\end{figure}

We model the code design proposal distribution as an uncorrelated multivariate Gaussian family as depicted in Fig.~\ref{fig:gauss_pg}, where each component corresponds to an index in the complete family of binary spreading codes. Our neural network model represents this proposal distribution by parameterizing the mean vector of the Gaussian distribution via a bounded output activation function, such as the logistic or hyperbolic tangent output activations. The variance~$\sigma^2$ of each component is maintained as a constant value. We can represent this Gaussian proposal distribution in terms of the network parameters~$\theta$ as
\begin{align}
p_\theta(x) &= \mathcal{N}\left(\mu(\theta), \sigma^2 \mathcal{I}^{(K\ell)} \right) \ , \label{pol_gauss}
\end{align}
where~$K$ denotes the number of codes in the family, $\ell$ denotes the length of each sequence in the code family, $\mathcal{I}^{(m)}$ represents the identity matrix of size~$(m \times m)$ where~$m$ is a positive integer, and $\mu(\theta)$ represents the mean of the Gaussian proposal distribution, as output by the GNN with parameters $\theta$. Note that the mean of~$p_\theta$ is bounded, i.e.~$\mu(\theta)\in [\mu_L, \mu_U]^{K\ell}$, due to the bounded nature of the output activation function of the network.

Due to the uncorrelated model of the multivariate Gaussian proposal distribution, each component represents a univariate Gaussian distribution over each binary value in the spreading code family. As depicted in Fig.~\ref{fig:gauss_pg}, to obtain a random output family of spreading codes, the code generator first samples directly from the multi-variate Gaussian proposal distribution defined in Eq.~\eqref{pol_gauss}, i.e.~$x \sim p_\theta$. From this sampled vector~$x$, the output code set is deterministically obtained by discretizing each component according to a threshold~$\zeta$ defined in Eq.~\eqref{def_midpoint}, which {we choose to be the midpoint} of the bounded output range of the GNN. Correspondingly, the output binary code family~$\overline{x}$ can be defined in an element-wise manner as in Eq.~\eqref{trans_fcn}:
\begin{align}
	x_i &\sim \mathcal{N}\left( \mu_i(\theta), \sigma^2 \right) \ . \label{pol_gauss_elem} \\
	\zeta &\coloneqq \frac{\mu_L + \mu_U}{2} \label{def_midpoint} \\
	\overline{x}_i &= \mathbbm{1}\{x_i \geq \zeta\} \ . \label{trans_fcn} 
\end{align}
Thus, after sampling the design point~$x$ from the proposal distribution~$p_\theta$, if a component of $x$ is above the threshold~$\zeta$, the corresponding sampled bit in the output binary code set~$\overline{x}$ is a~$1$; otherwise, the corresponding bit is a~$0$. Note that the design space~$\mathcal{X}$, where~$x \in \mathcal{X}$, is in fact a continuous design space. Indeed, we map each continuous design point~$x$ to a binary vector~$\overline{x}$ according to the threshold~$\zeta$ as in Eq.~\eqref{trans_fcn}. Correspondingly, we evaluate the performance of the design point~$x$ by utilizing its discretized binary vector~$\overline{x}$ with the spreading code evaluation metric, as defined in Sec.~\ref{ssec:metrics}.

\subsection{Spreading Code Evaluation Metric} \label{ssec:metrics}
In the NES framework, the evaluation metric provides feedback to the code generator in order to improve its proposal distribution. In this work, we consider the \textit{mean-square non-central even auto-correlation} and the \textit{mean-square even cross-correlation} as the two objectives to minimize, defined respectively as
\begin{align}
	f_{AC} &\coloneqq \frac{1}{K\ell} \left( \sum_{k=1}^K \sum_{\delta=1}^{\ell-1} |R_k(\delta)|^2 \right) \label{def_auto_comp} \\
	f_{CC} &\coloneqq \frac{1}{K_p \ell} \left( \sum_{k=1}^K \sum_{j=k+1}^K \sum_{\delta=0}^{\ell-1} |R_{k,j}(\delta)|^2 \right) \ , \label{def_cross_comp}
\end{align}
where~$K$ represents the number of sequences in the output code family~$\overline{x}$, $\ell$ represents the length of each sequence, $K_p \coloneqq \begin{pmatrix} K \\ 2 \end{pmatrix}$ represents the number of pairs of sequences in the code family of size~$K$, $R_k$ represents the auto-correlation of the $k^{th}$ sequence in the output code family~$\overline{x}$ as defined in Eq.~\eqref{eq:auto_corr_def}, and $R_{k,j}$ represents the cross-correlation of sequences $k$ and $j$ of the output code family~$\overline{x}$ as defined in Eq.~\eqref{eq:cross_corr_def}. 

In order to ensure the code generator reduces both objectives, we define the comprehensive objective function as the maximum of the two components from Eqs.~\eqref{def_auto_comp}~and~\eqref{def_cross_comp}
\begin{align}
	f(\overline{x}) &= \max\left(f_{AC}, f_{CC} \right) \ . \label{reward_def}
\end{align}
Thus, to minimize the objective, the NES algorithm will seek to reduce both components of the objectives simultaneously, without compromising one component over the other. Indeed, modifications of the objective function in Eq.~\eqref{reward_def} could be readily incorporated within this machine learning framework. In particular, if seeking to design spreading codes with smaller auto-correlation sidelobes than cross-correlation sidelobes, one could scale $f_{AC}$ by a corresponding factor that is greater than $1$ in Eq.~\eqref{reward_def}, which would encourage the algorithm to generate codes that reduce the auto-correlation metric in comparison to the cross-correlation metric by the corresponding user-defined factor. 

Furthermore, additional components of objectives could be incorporated into the comprehensive objective function, including, for example, odd correlation performance metrics and relative Doppler frequency correlation metrics. These additional objectives could be incorporated as new arguments within the maximization function of Eq.~\eqref{reward_def} with relative scaling factors if desired, or these objective could be combined via a weighted summation, as proposed in~{\citep{soualle2005spreading}}. 
	
\subsection{NES Optimization}
We optimize the network parameters~$\theta$, which define the proposal distribution~$p_\theta$, by following the NES optimization process described in Section~\ref{ssec:NES}. To reduce the variance in the gradient estimate, we incorporate a constant baseline~$b$, which is subtracted from the objective function in the gradient estimate expression of Eq.~\eqref{eq:grad_est}~\citep{mohamed2020monte}. Thus, with the baseline subtraction, the gradient estimate from Eq.~\eqref{eq:grad_est} becomes 
\begin{align}
	\nabla_\theta J(\theta) &\approx \frac{1}{N} \sum_{i=1}^N  \left( f(x^i) - b \right) \nabla_\theta \log p_\theta(x^i) \ , \label{grad_est_def_wBL}
\end{align}
where~$N$ represents the cardinality of the batch of samples~$\{x^1, x^2, \ldots, x^N\}$. For our constant baseline~$b$, we utilize the mean return in the current batch of samples, defined as
\begin{align}
	b &\coloneqq \frac{1}{N} \sum_{i=1}^N f(x^i) \ . \label{BL_def}
\end{align} 
Subtracting a constant value does not introduce any bias in the gradient estimate, but it can significantly reduce its variance~\citep{wierstra2008natural,mohamed2020monte}. Indeed, by reducing the variance of the gradient estimate, we observe improved convergence and overall learning performance during training. {Section~\ref{ssec:baseline} demonstrates the learning rate of the NES algorithm with and without the incorporation of this baseline from Eq.~\eqref{BL_def}.}

%% file: sections/experiments.tex
\section{Experimental Validation} \label{sec:exp_val}

\subsection{Details of Experimental Setup} \label{ssec:exp_setup}
We validate the ability of our NES machine learning algorithm to devise low-correlation spreading codes and further compare its performance with that of well-chosen families of equal-length Gold codes and Weil codes. We compare the performance of our algorithm with Gold codes of length-$63$, $127$, $511$, and $1023$ bits. Similarly, since Weil codes only exist for sequence lengths that correspond to a prime number length, we compare our algorithm with Weil codes of length-$67$, $127$, $257$, $521$, and $1031$ bits.

\begin{table}
\renewcommand{\arraystretch}{1.3}
\caption{Design parameters of proposed NES machine learning method.}
\label{tab:pg_params}
\centering
\begin{tabular}{|c|c|}
\hline
\textbf{Parameter Description} & \textbf{Value} \\
\hline
Gaussian proposal variance $(\sigma^2)$ & $0.1$ \\
\hline
learning rate (for $\ell < 500$) & $10^{-4}$ \\
\hline
learning rate (for $\ell > 500$) & $5\cdot 10^{-5}$ \\
\hline
hidden layer size & $2 K \ell$ \\
\hline
number hidden layers & $2$ \\
\hline
output activation & \texttt{tanh} \\
\hline
network optimizer & \texttt{Adam}~\citep{kingma2014adam} \\
\hline
batch size $(N)$ & $100$ \\
\hline
number iterations & $10,000$ \\
\hline
\end{tabular}
\end{table}

From our NES method, we generate sequences for family sizes of $3$~codes up to $31$~codes. We additionally compare our proposed algorithm with the \textit{best performing} code family across $10,000$~sampled families of Gold codes and Weil codes. In a few of the sample runs of Gold and Weil codes, we observed a large deviation in the auto- and cross-correlation cost components which frequently leads to worse performance on the overall objective defined in Eq.~\eqref{reward_def}. In these instances, {when possible to reduce the deviation in auto- and cross-correlation objectives}, we would resample the conventional code families, leading to an improvement in the performance metric of the Gold and Weil codes.
{The Gold and Weil code performance results have been updated since those presented in our previous conference paper~\citep{mina2020designing}, resulting in similar or slightly worse performance of the Gold and Weil codes on the overall objective.}

The details of our NES machine learning algorithm are indicated in Tab.~\ref{tab:pg_params}. We utilize a slightly smaller learning rate for learning families with sequences of over length-500 bits, since we observed more consistent performance with a reduced learning rate for these longer sequence test cases. We use the hyperbolic tangent function as our bounded output activation function and Adam~\citep{kingma2014adam} as the optimizer of the GNN.

Furthermore, we compare the performance of our algorithm with that of an analogous genetic algorithm which is set to optimize over the same maximization evaluation objective defined in Eq.~\eqref{reward_def}. The GA implementation uses \textit{fitness proportionate selection}~\citep{holland1975adaptation}, which stochastically selects a candidate solution from the population with a probability proportional to its normalized objective. Selected individuals are recombined using \textit{uniform crossover}~\citep{syswerda1989uniform} which allows for greater recombination of candidate solutions and was observed to have better performance during initial testing. We additionally incorporated \textit{elitism}~\citep{baluja1995removing}, which ensures that the genetic algorithm will continuously improve during its optimization process. 
\begin{table}
	\renewcommand{\arraystretch}{1.3}
	\caption{Design parameters of genetic algorithm implementation.}
	\centering
	\begin{tabular}{|c|c|}
		\hline
		\textbf{Parameter Description} & \textbf{Value} \\
		\hline
		elite rate & $0.01$ \\
		\hline
		mutation rate & $0.005$ \\
		\hline
		selection method & \texttt{fitness proportionate}~\citep{holland1975adaptation} \\
		\hline
		crossover method & \texttt{uniform}~\citep{syswerda1989uniform} \\
		\hline
		population size & $100$ \\
		\hline
		number iterations & $10,000$ \\
		\hline
	\end{tabular}
	\label{tab:ga_params}
\end{table}

In Tab.~\ref{tab:ga_params}, we  describe the parameters of the analogous genetic algorithm. For the GA to be comparable with the NES method, we set its population size to be the batch size of the NES machine learning algorithm, and we set the number of iterations for both algorithms to be the same.

\subsection{Effect of Baseline Incorporation on Learning Performance} \label{ssec:baseline}
\begin{figure}
	\centering
	\begin{subfigure}{0.45\textwidth}
		\centering
		\includegraphics[width=\textwidth]{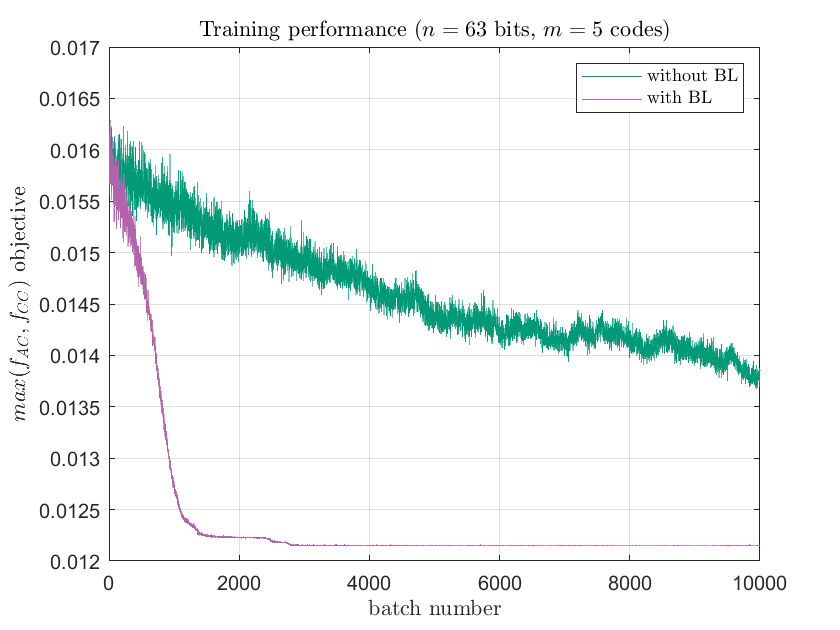}
		\caption{$63$ bits, $5$ codes}
	\end{subfigure}
	\begin{subfigure}{0.45\textwidth}
		\centering
		\includegraphics[width=\textwidth]{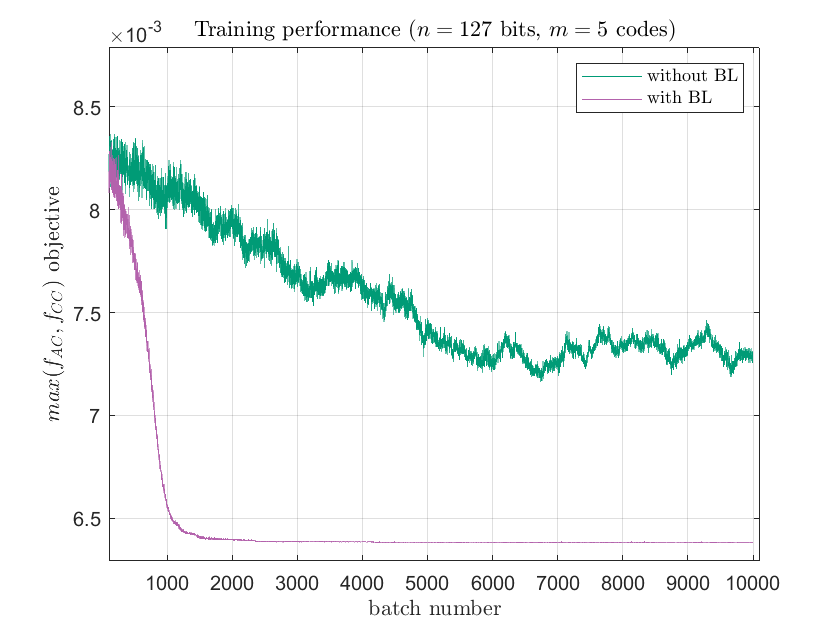}
		\caption{$127$ bits, $5$ codes}
	\end{subfigure}
	\caption{Training performance on the normalized maximum correlation objective, i.e.~$\max(f_{AC}, f_{CC})$, of the NES machine learning algorithm, with (violet) and without (green) baseline incorporation. We observe that baseline incorporation leads to faster and more consistent learning performance during training.}
	\label{fig:baseline_incorp}
\end{figure}

Fig.~\ref{fig:baseline_incorp} demonstrates the effect of incorporating the constant mean baseline on the learning performance during training, as measured by the normalized maximum correlation objective, i.e.~$\max\{f_{AC}, f_{CC}\}$, which the algorithm seeks to minimize. For both of the code length scenarios shown in Fig.~\ref{fig:baseline_incorp}, we observe that the incorporation of the baseline, shown in violet, leads to significant improvement in the learning performance during training, due to the reduction of variance in the gradient estimate. In particular, the baseline incorporation allows for more systematic learning and more consistent improvement on the correlation objective.
	
\subsection{Comparison with Best Performing Gold and Weil Codes}
\begin{figure}
	\centering
	\begin{subfigure}{0.4\textwidth}
		\centering
		\includegraphics[width=\textwidth]{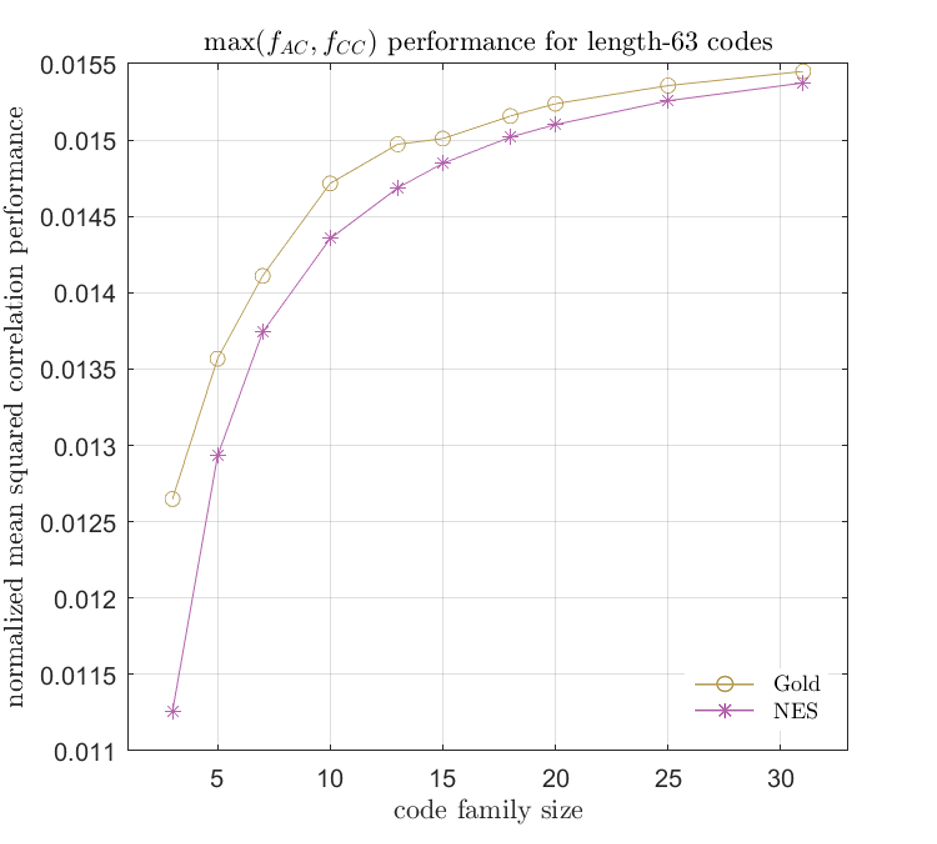}
		\caption{length-$63$ codes}
	\end{subfigure}
	\begin{subfigure}{0.4\textwidth}
		\centering
		\includegraphics[width=\textwidth]{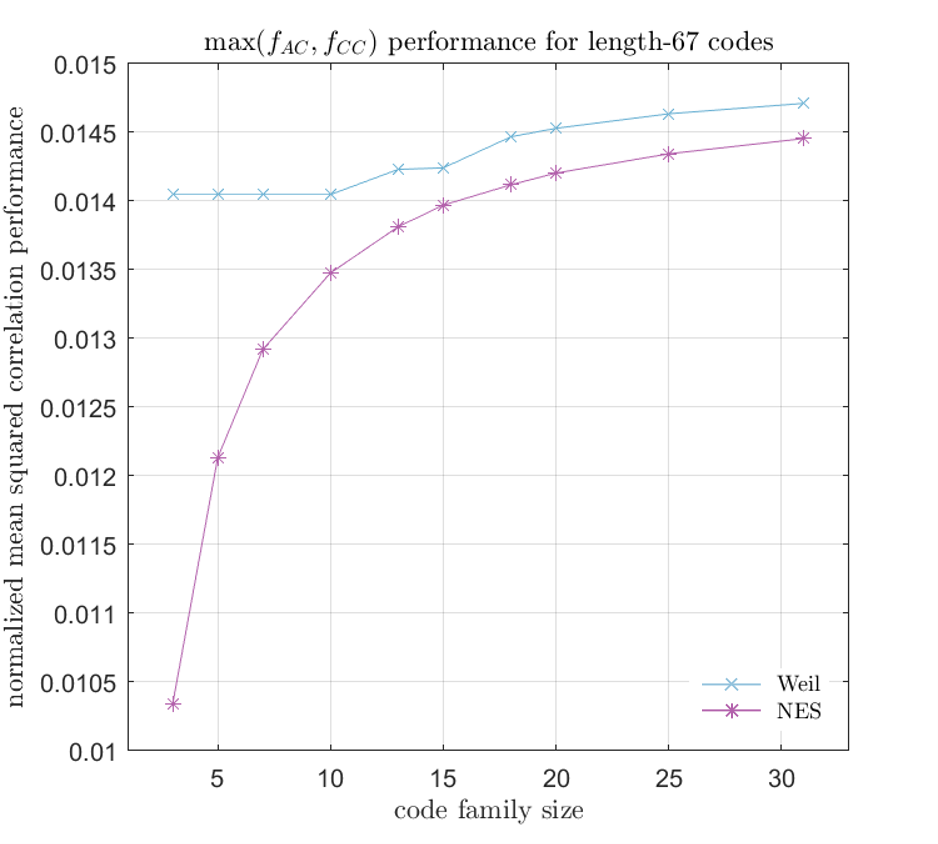}
		\caption{length-$67$ codes}
	\end{subfigure}
	\begin{subfigure}{0.4\textwidth}
		\centering
		\includegraphics[width=\textwidth]{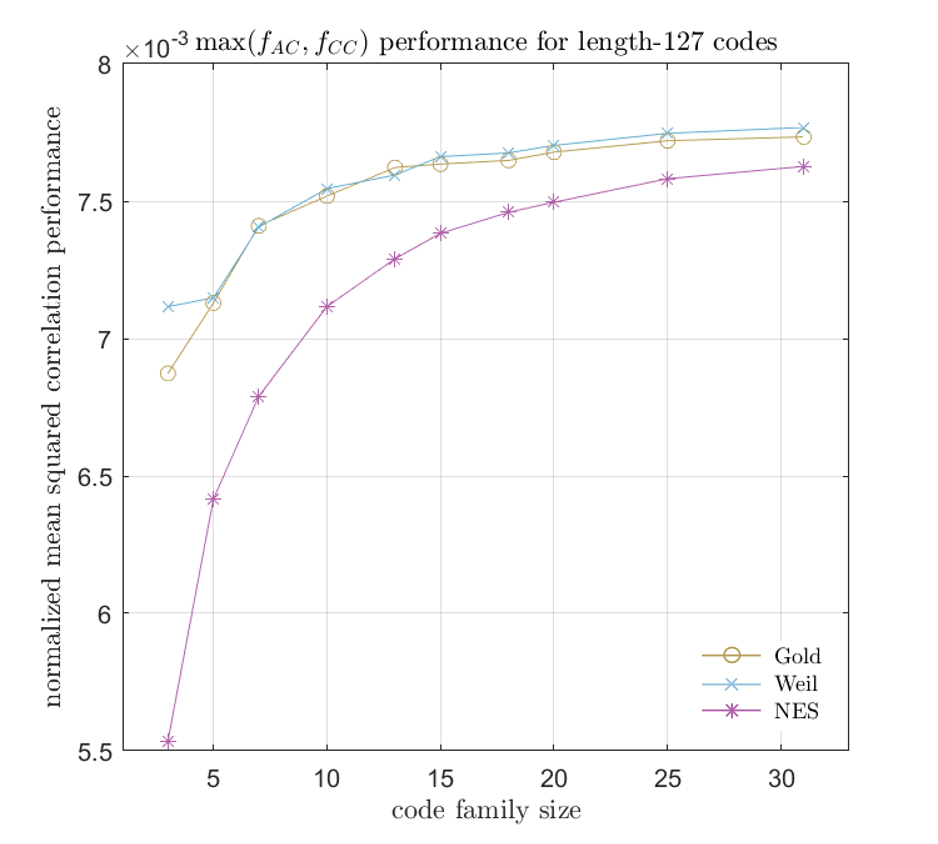}
		\caption{length-$127$ codes}
	\end{subfigure}
	\begin{subfigure}{0.4\textwidth}
		\centering
		\includegraphics[width=\textwidth]{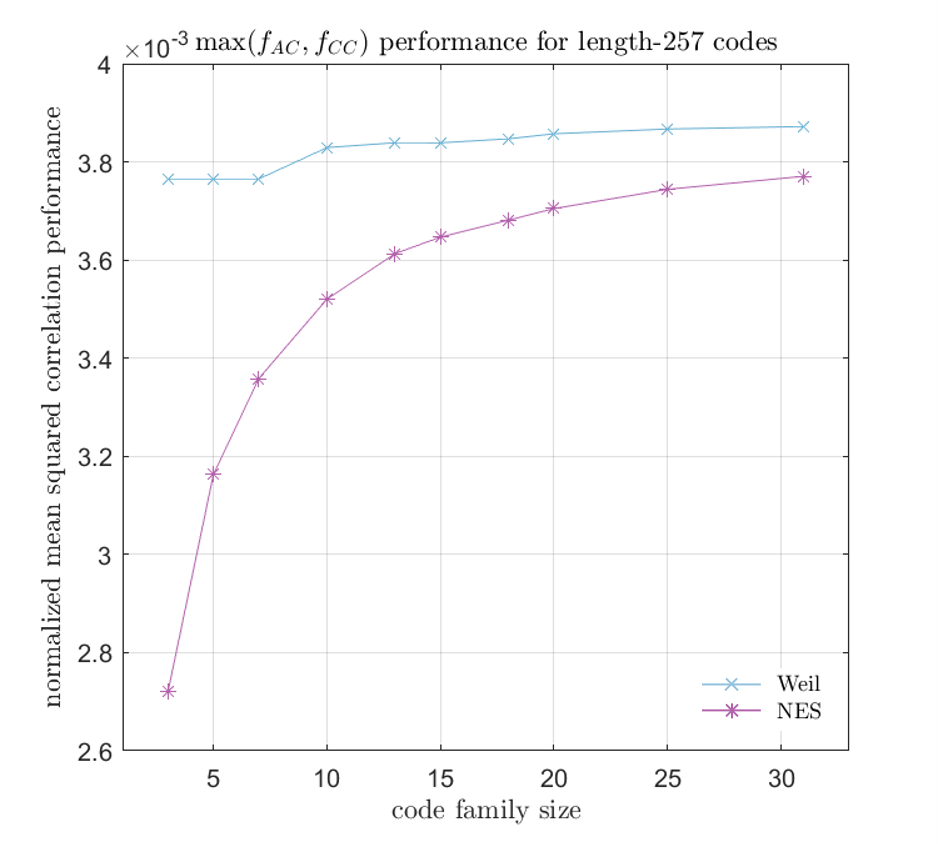}
		\caption{length-$257$ codes}
	\end{subfigure}
	\begin{subfigure}{0.4\textwidth}
		\centering
		\includegraphics[width=\textwidth]{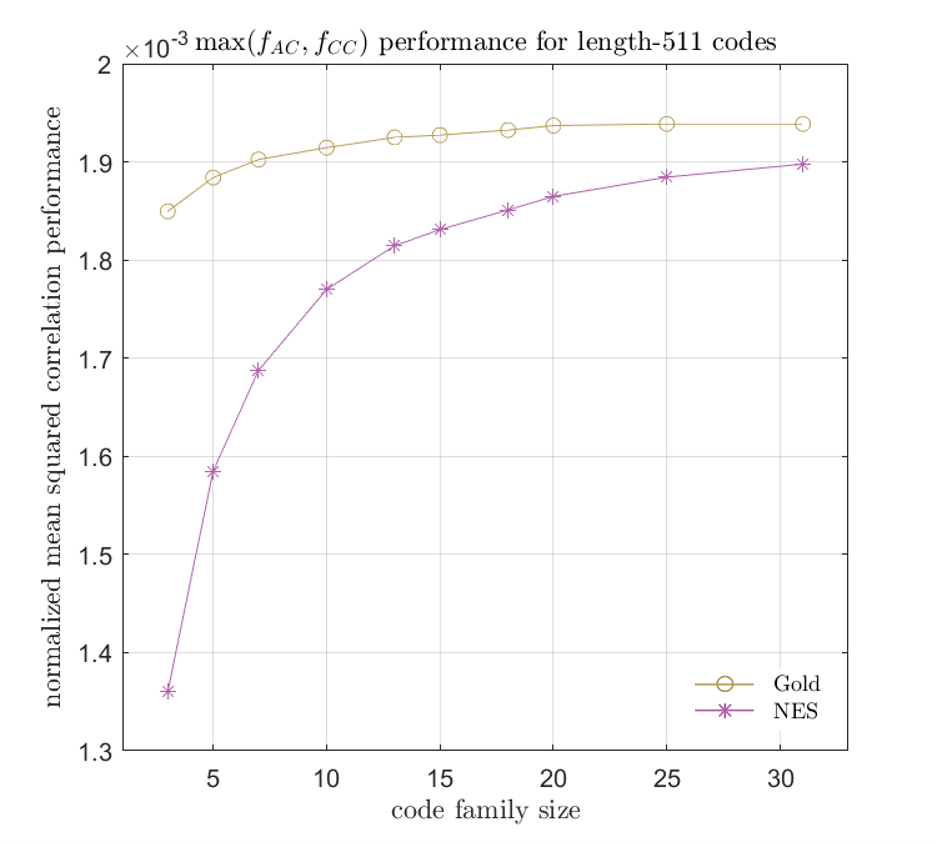}
		\caption{length-$511$ codes}
	\end{subfigure}
	\begin{subfigure}{0.4\textwidth}
		\centering
		\includegraphics[width=\textwidth]{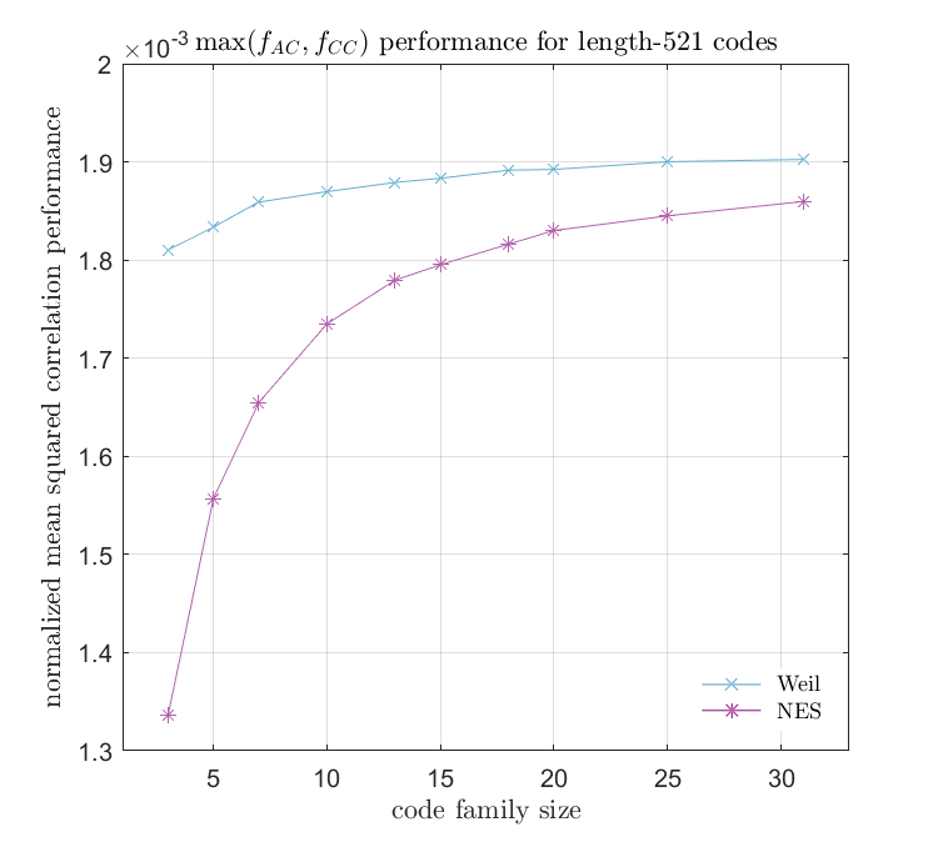}
		\caption{length-$521$ codes}
	\end{subfigure}
	\caption{{Comparison of the proposed NES machine learning method with that of well-chosen Gold and Weil code families as a function of family size. Performance is optimized in terms of the maximization evaluation objective, i.e. $\max(f_{AC}, f_{CC})$, defined in Eq.~\eqref{reward_def}, plotted here in terms of the normalized correlations. The Gold code correlation performance is indicated in the gold line with circle markers denoting the particular, discrete set of family sizes tested in each case, while the Weil code performance is indicated in blue with $\times$ markers and the performance of the codes generated by our proposed NES algorithm is indicated in violet with asterisk markers. We observe that the NES method outperforms the best Gold and Weil codes across all tested sequence lengths and code family sizes.}}
	\label{fig:gold_weil}
\end{figure}

\begin{figure}
	\centering
	\begin{subfigure}{0.4\textwidth}
		\centering
		\includegraphics[width=\textwidth]{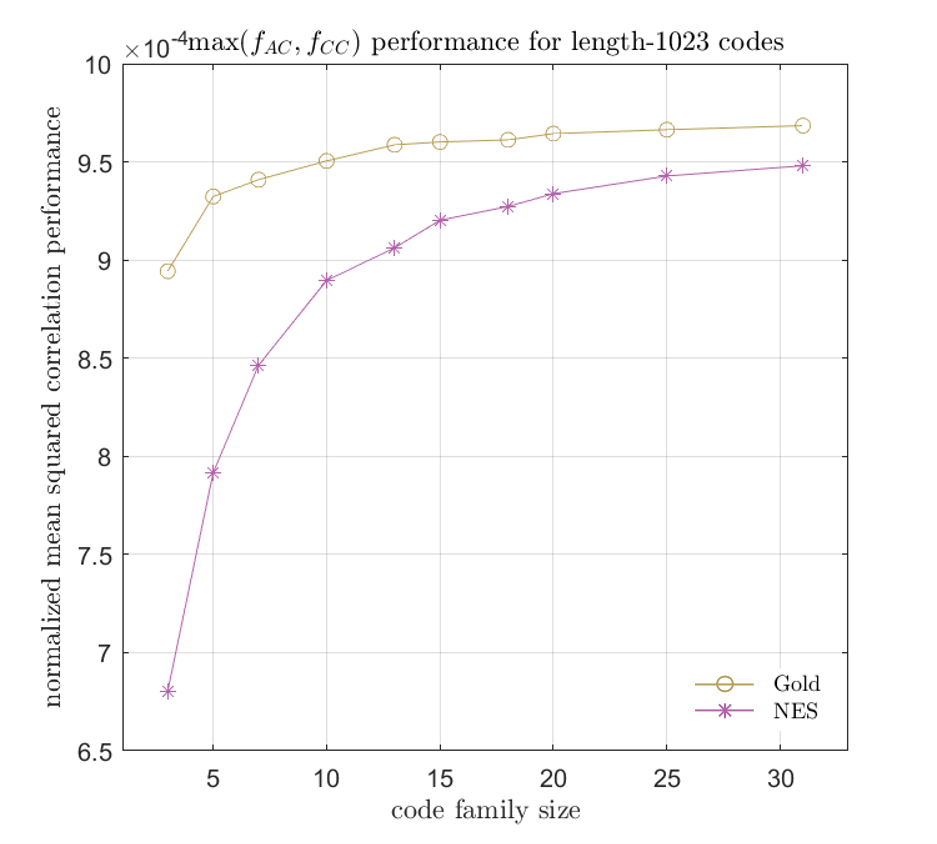}
		\caption{length-$1023$ codes}
	\end{subfigure}
	\begin{subfigure}{0.4\textwidth}
		\centering
		\includegraphics[width=\textwidth]{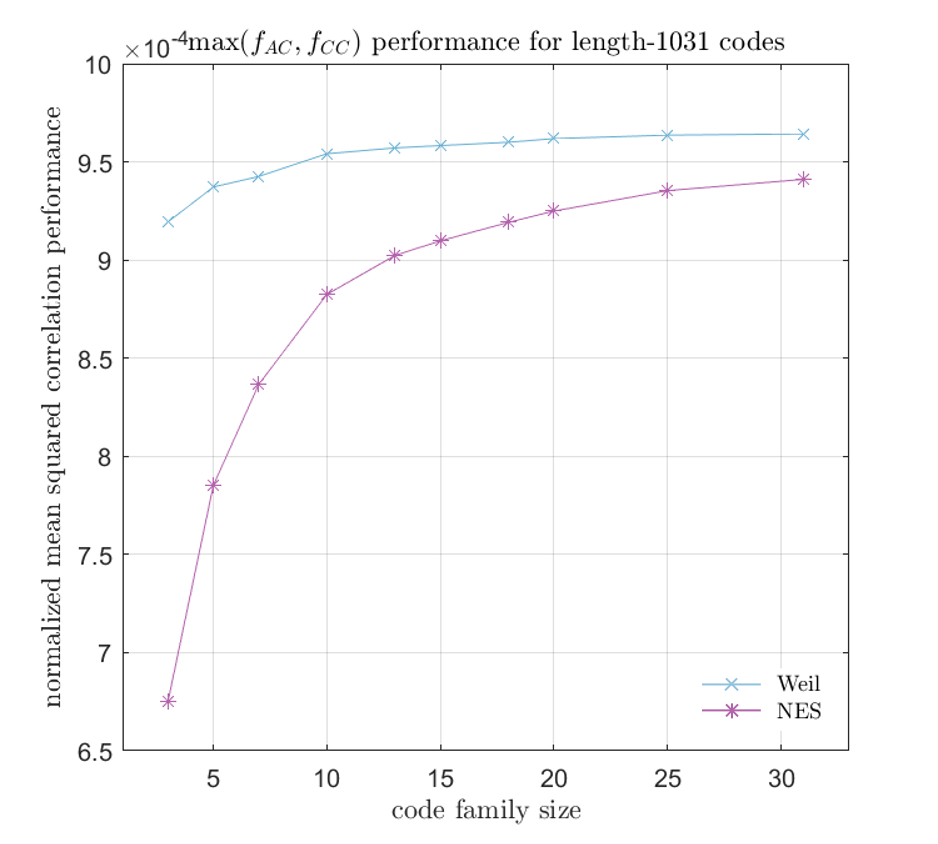}
		\caption{length-$1031$ codes}
	\end{subfigure}
	\caption{{Comparison of the proposed NES machine learning method with that of well-chosen Gold and Weil code families as a function of family size for length-1023 and length-1031 sequences. Performance is optimized in terms of the maximization evaluation objective, i.e. $\max(f_{AC}, f_{CC})$, defined in Eq.~\eqref{reward_def}, plotted here in terms of the normalized correlations. The Gold code correlation performance is indicated in the gold line with circle markers denoting the particular, discrete set of family sizes tested in each case, while the Weil code performance is indicated in blue with $\times$ markers and the performance of the codes generated by our proposed NES algorithm is indicated in violet with asterisk markers. We observe that the NES method outperforms the best Gold and Weil codes across all tested code family sizes.}}
	\label{fig:gold_weil_2}
\end{figure}

{Figs.~\ref{fig:gold_weil}~and~\ref{fig:gold_weil_2} show the converged performance of our NES algorithm after training, comparing it with the best-performing sets of Gold and Weil code families of equal-length. 
We plot the final performance in terms of the maximization evaluation objective defined in Eq.~\eqref{reward_def}, i.e.~$\max\{f_{AC}, f_{CC}\}$, as a function of the code family size.
From these results, we empirically observe how designing spreading code sequences for larger family sizes generally leads to worse performance on the optimization objective, for all types of code sequences.
As a result, developing a flexible optimization framework to tailor the design of the code family to the desired, smaller number of spreading codes may lead to better performance of the complete navigation system.}

{From Figs.~\ref{fig:gold_weil}~and~\ref{fig:gold_weil_2}, we observe that the proposed NES method in violet outperforms the best Gold code and Weil code families across all tested code lengths and code family sizes.
To determine if this performance improvement against the best-performing sets of Gold and Weil codes continues for larger families of the NES-generated spreading codes, further testing would be required with more extensive memory and computational resources. 
In Sec.~\ref{ssec:scaling}, we discuss how the proposed NES method scales with the spreading code length and the size of the code family, while discussing potential techniques to reduce the memory requirements and resources during the training process.}

\subsection{Comparison with Analogous Genetic Algorithm + Elitism}
\begin{figure}
	\centering
	\begin{subfigure}{0.4\textwidth}
		\centering
		\includegraphics[width=\textwidth]{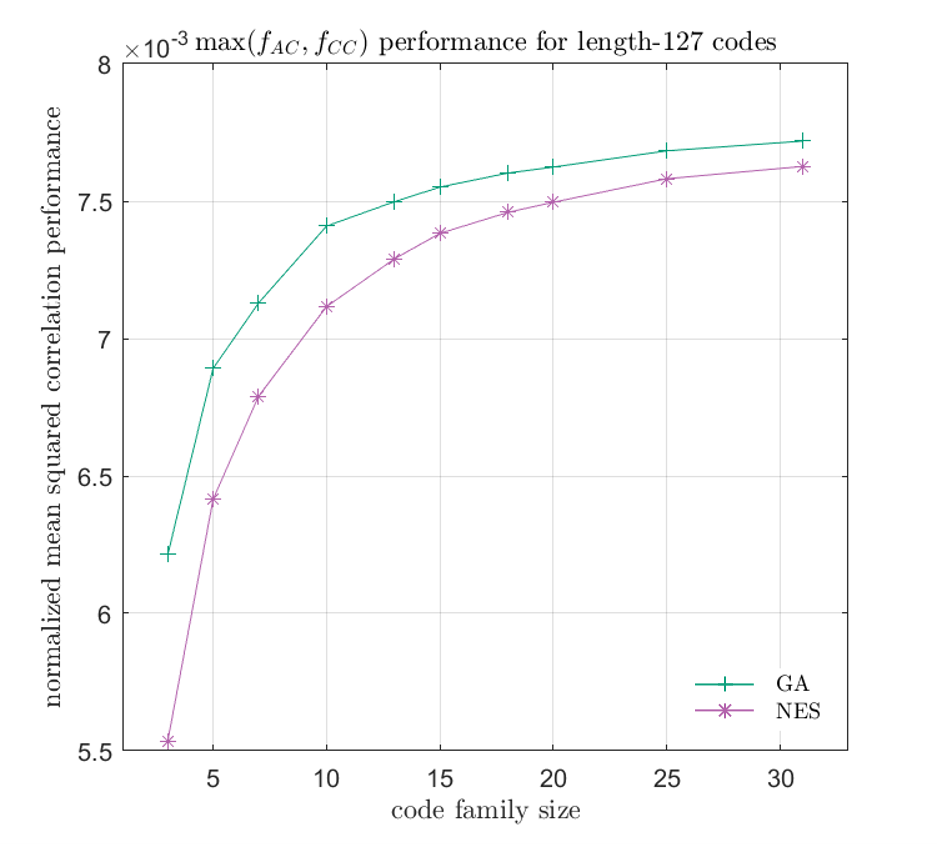}
		\caption{length-$127$ codes}
	\end{subfigure}
	\begin{subfigure}{0.4\textwidth}
		\centering
		\includegraphics[width=\textwidth]{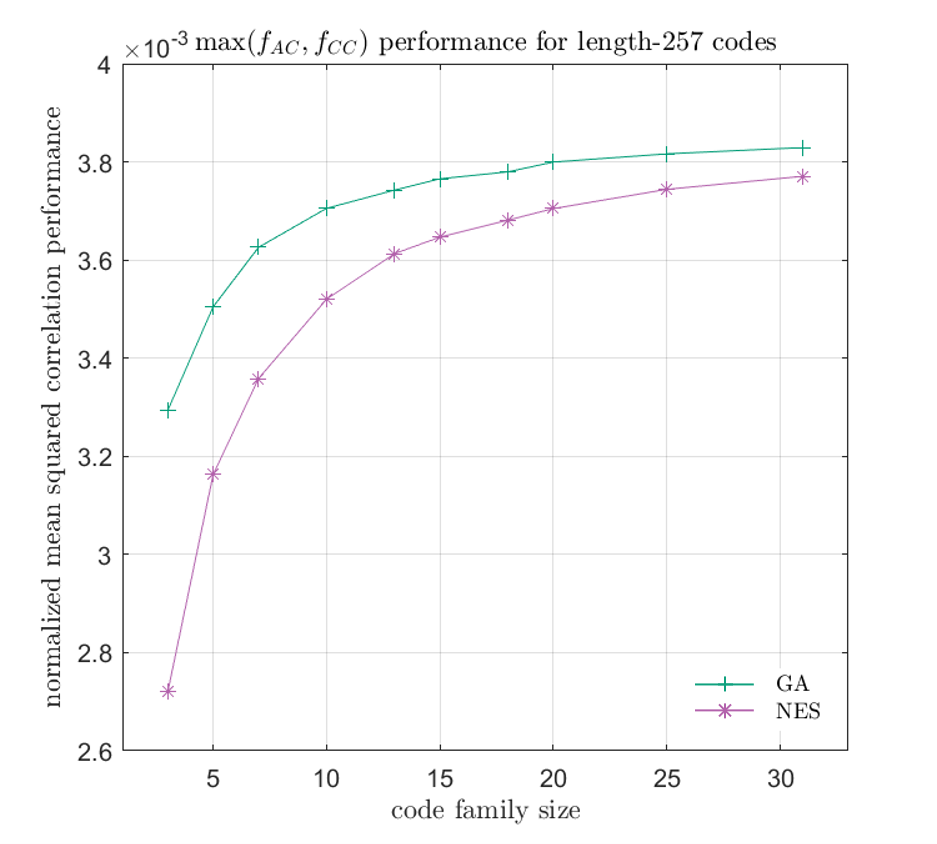}
		\caption{length-$257$ codes}
	\end{subfigure}
	\begin{subfigure}{0.4\textwidth}
		\centering
		\includegraphics[width=\textwidth]{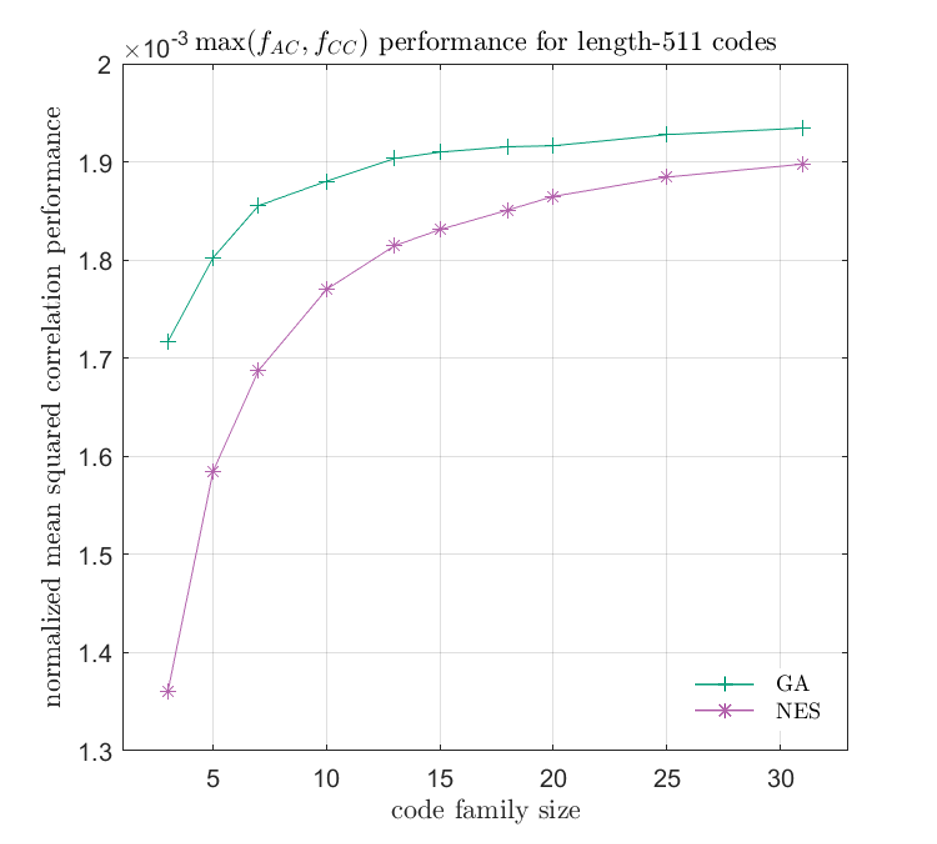}
		\caption{length-$511$ codes}
	\end{subfigure}
		\begin{subfigure}{0.4\textwidth}
		\centering
		\includegraphics[width=\textwidth]{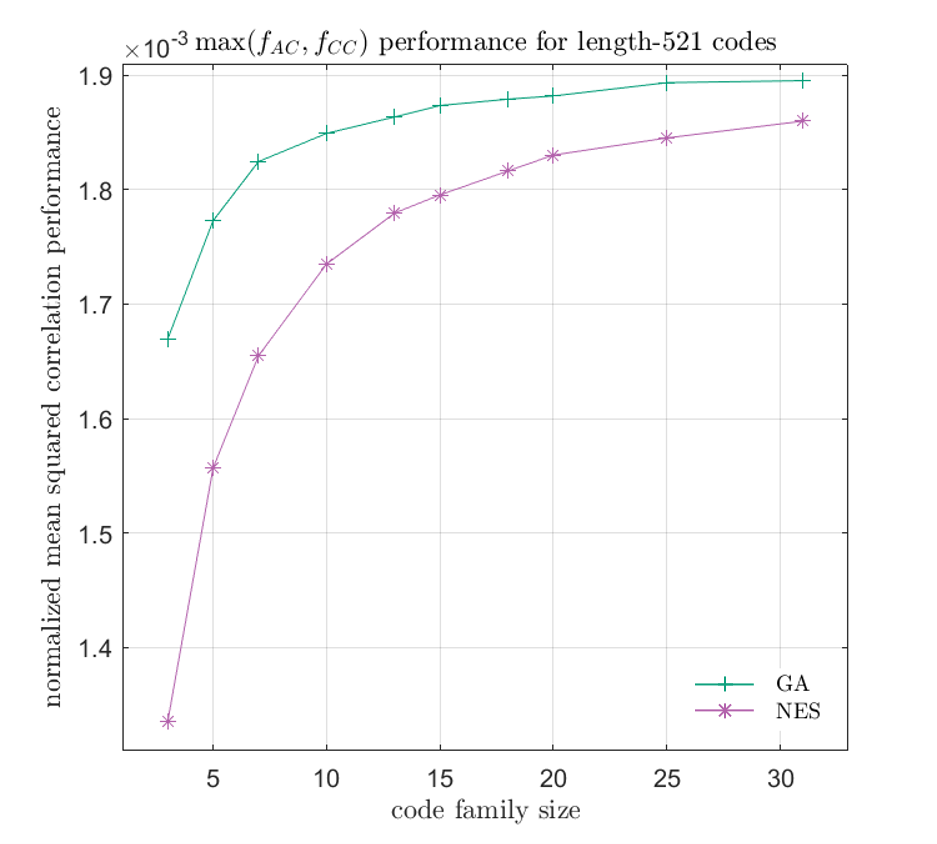}
		\caption{length-$521$ codes}
	\end{subfigure}
	\begin{subfigure}{0.4\textwidth}
		\centering
		\includegraphics[width=\textwidth]{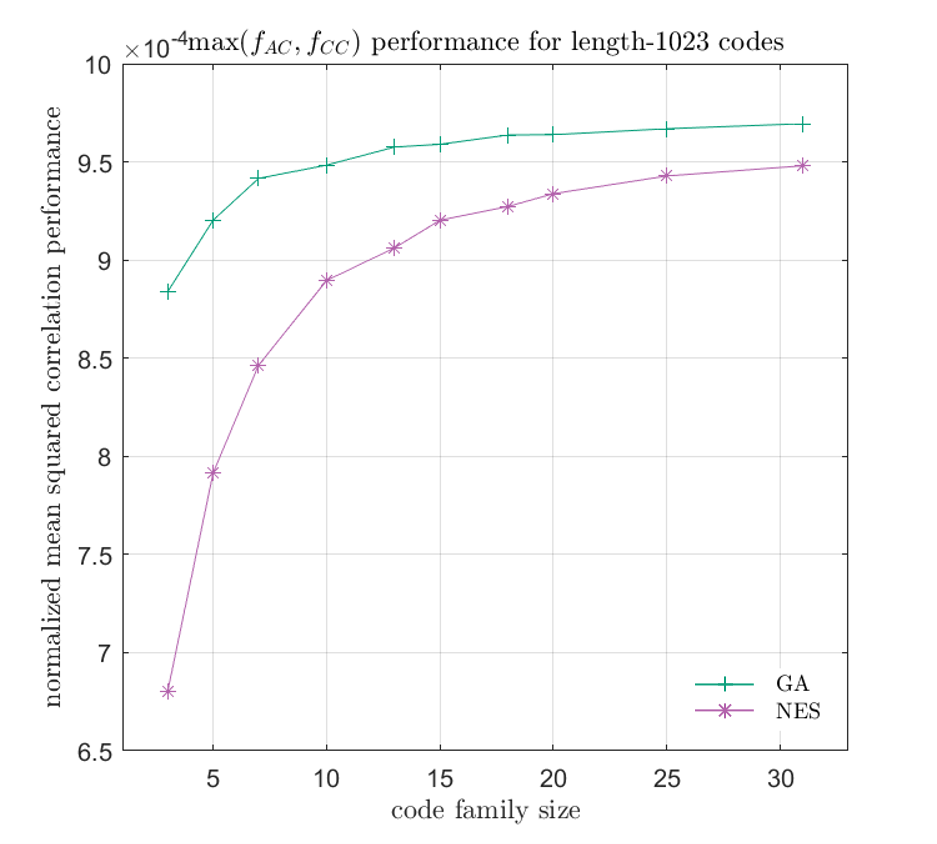}
		\caption{length-$1023$ codes}
	\end{subfigure}
	\begin{subfigure}{0.4\textwidth}
		\centering
		\includegraphics[width=\textwidth]{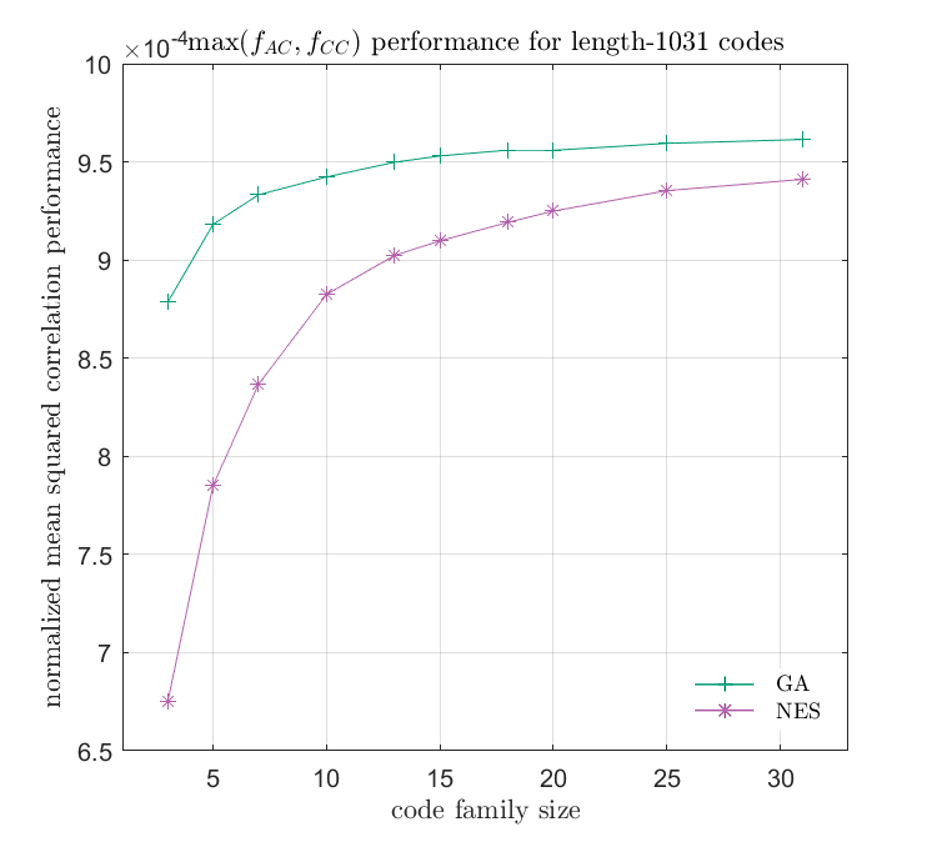}
		\caption{length-$1031$ codes}
	\end{subfigure}
	\caption{{Comparison of the proposed NES machine learning method with that of the analogous genetic algorithm with incorporated elitism. Performance is optimized in terms of the maximization evaluation objective, i.e. $\max(f_{AC}, f_{CC})$, defined in Eq.~\eqref{reward_def}, plotted here in terms of the normalized correlations. The genetic algorithm (GA) performance is indicated in green with $+$ markers denoting the particular, discrete set of family sizes tested in each case, while the performance of our proposed NES method is indicated in violet with asterisk markers. We observe that for all spreading code lengths, the NES method outperforms the GA implementation in designing spreading codes which minimize the spreading code evaluation metric in Eq.~\eqref{reward_def}.}}
	\label{fig:ga_perf}
\end{figure}

{Finally, we compare our NES method with that of the analogous genetic algorithm described in Sec.~\ref{ssec:exp_setup}. 
Fig.~\ref{fig:ga_perf} shows separate plots of the normalized correlation performance from length-127 up to length-1031 bit spreading code families. 
Both optimization algorithms sought to minimize the same spreading code correlation objective defined in Eq.~\eqref{reward_def}.
The GA additionally incorporates elitism to guarantee the algorithm continuously improves during its optimization process.
We observe in Fig.~\ref{fig:ga_perf} that our NES method consistently outperforms the implemented GA with elitism across all tested sequence lengths and family sizes.}

\subsection{Scaling of Network Size with the Size of the Learned Code Family} \label{ssec:scaling}
With this proposed method, the total number of parameters in the GNN, and correspondingly the memory required for the network to train, scales quadratically with the length of the total code set to be determined, i.e.~$K\ell$ total binary values for length-$\ell$ binary sequences and $K$~codes. Indeed, these network parameters are only necessary during the training phase of the algorithm, while the codes are being learned. Once the algorithm converges to a final, optimized spreading code family, the network parameters are no longer needed and can safely be discarded. However, the memory required during training correspondingly scales quadratically with the length of the total code set, thereby requiring significantly greater investments in computing resources, especially for very large family sizes of longer spreading codes, such as a family of 420 length-10230 sequences as used for GPS~L1C~\citep{interface2021gpsl1c}.

Besides simplifying the neural network structure, one could reduce memory requirements for training large-scale neural networks by modeling network parameters with lower precision as well as by exploring the use of sparsity in the network model, where some of the network parameters are set to zero and are correspondingly no longer necessary to store in memory. In this work, we used a fully connected network model; however, leveraging sparsity has been shown to significantly reduce the number of network parameters required for a deep networks while achieving similar or even better performance~{\citep{sohoni2019low,mostafa2019parameter}}. Another advantage to utilizing sparse models is the corresponding reduction in the number of gradient computations as well as the size of the momentum buffer for the optimizer, thereby also leading to reduced memory requirements during training.

%% file: sections/conclusion.tex
\section{Conclusion} \label{sec:conc}
In this work, we developed a {flexible} framework to devise a family of high-performing spreading code sequences which achieve low mean-square periodic auto- and cross-correlation objectives. We utilize a natural evolution strategy algorithm using a generative neural network to model a Gaussian proposal distribution over the space of binary spreading code families. Using the Monte Carlo-based gradient estimate, we directly optimize the proposal distribution over the design search space. Furthermore, we believe this is the first work to develop a machine learning method to design navigation spreading code families. 

{By minimizing the maximum between the mean-squared auto-correlation and the mean-squared cross-correlation, we demonstrated the ability of our algorithm to achieve better performance than well-chosen families of Gold and Weil codes, for sequences of up to length-1023 and length-1031 bits and family sizes of up to 31 codes.
We additionally demonstrated that the spreading code families generated from our method consistently outperformed those obtained from a genetic algorithm with incorporated elitism on the overall objective function.}

%% file: sections/appendix.tex
\section{Appendix: Background on Neural Networks} \label{sec:app}
At a high-level, an \textit{artificial neural network} is a collection of parameterized linear and nonlinear functions, which altogether represent a complex and flexible mathematical model~\citep{bishop2006pattern}. The parameters of the network are optimized by the machine learning algorithm, or ``learned,'' in order to recognize potentially complicated patterns which are difficult to describe within a provided set of data.   
\begin{figure}
	\centering
	\includegraphics[width=0.6\textwidth]{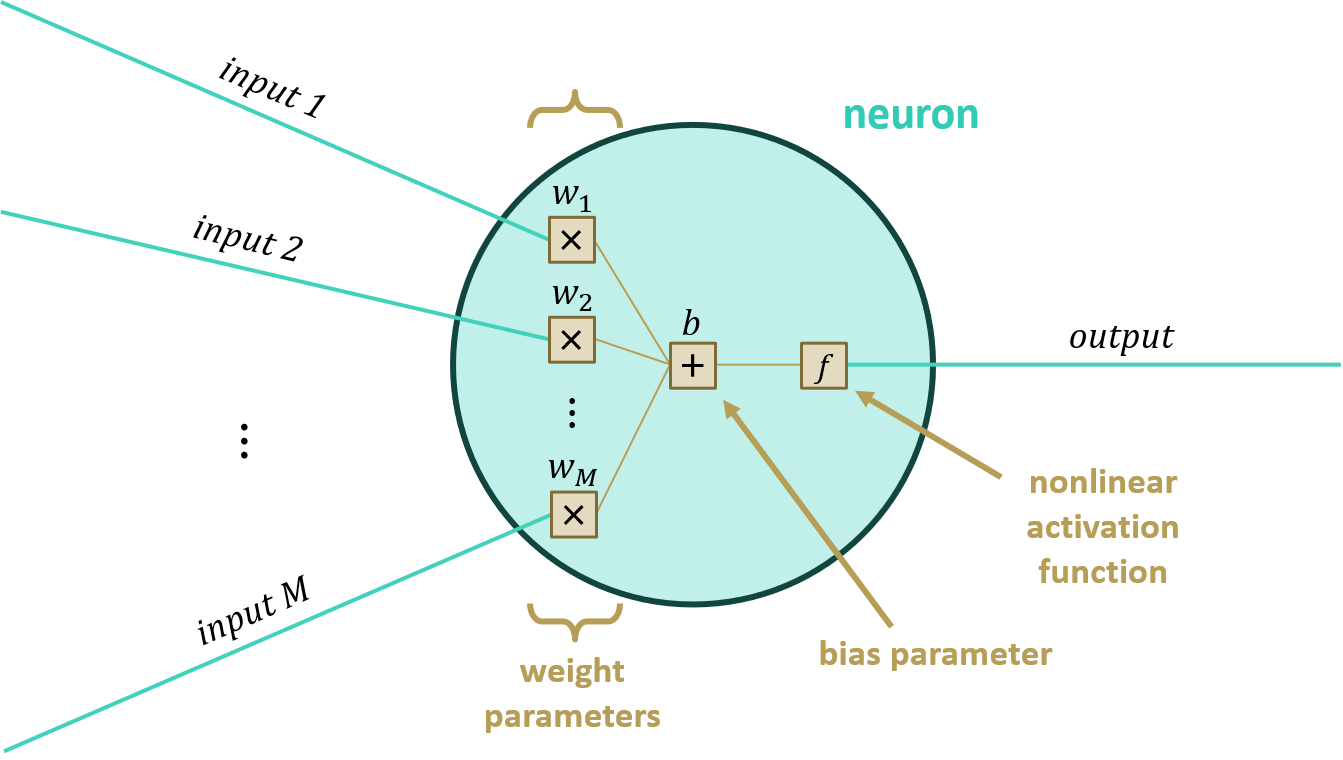}
	\caption{Depiction of an individual artificial neuron, the key structural component of an artificial neural network.}
	\label{fig:neuron}
\end{figure}

The key structural component of the neural network architecture is an \textit{artificial neuron}, which is pictorally represented in Fig.~\ref{fig:neuron}. A neuron consists of three main components: (1) \textit{weight parameters}, (2) a \textit{bias parameter}, and a (3) \textit{nonlinear activation} function as shown in Fig.~\ref{fig:neuron}. A neuron can have any number of inputs; in Fig.~\ref{fig:neuron}, we illustrate a neuron with $M$~inputs. Each input~$i \in \{1, \ldots, M\}$ is multiplied by its corresponding weight~$w_i$ which is a parameter of the neuron, then the weighted inputs are summed and added to the bias term~$b$, which is also a parameter of the network. Finally, the summed term, which is a scalar value for a particular neuron, is input to a nonlinear activation function represented as~$f$. Indeed, the notation~$f(\cdot)$ is commonly used in machine learning to represent the nonlinear activation function of a neuron, and in this case should not be confused with the objective function used for optimization. Thus, the explicit expression for the mathematical operations performed by an individual neuron is
\begin{align}
	y &= f\left( b + \sum_{i=1}^M x_i w_i \right), 
\end{align}
where $y$ is the scalar value output by the neuron and $x_i$ is the $i^{th}$~input value to the neuron.

\begin{figure}
	\centering
	\includegraphics[width=0.3\textwidth]{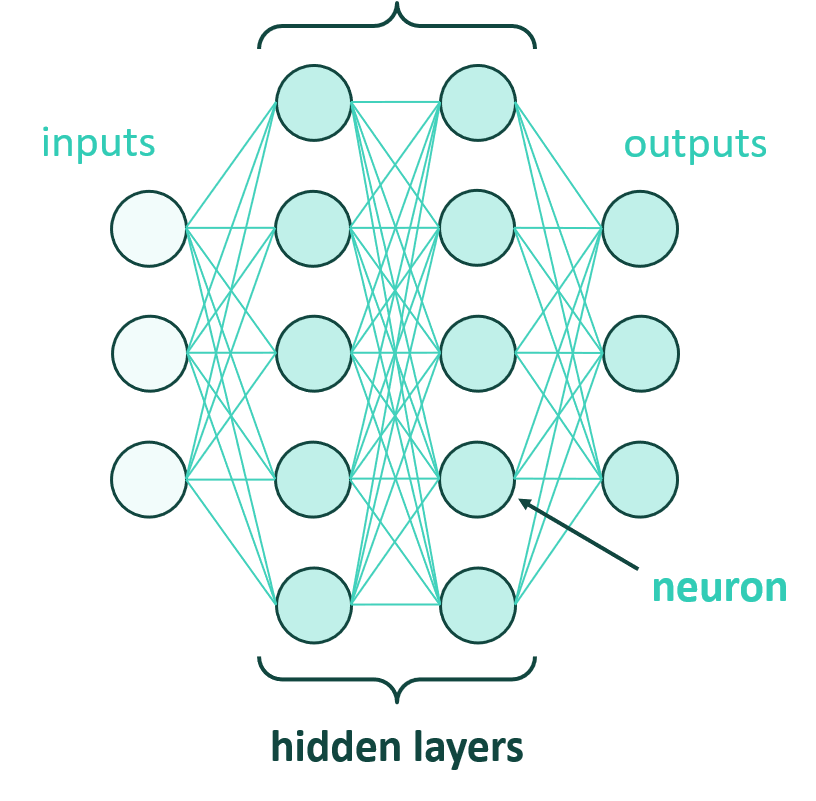}
	\caption{Depiction of an artificial neural network, consisting of layers of ``stacked'' artificial neurons.}
	\label{fig:neural_net}
\end{figure}

Nonlinear activation functions have a domain that is continuous and unbounded, i.e.~$\mathbb{R}$, and generally a range that is continuous, which may or may not be bounded. Common bounded-output activation functions include the \textit{logistic sigmoid} function, commonly called a \textit{sigmoid} function, as defined in Eq.~\eqref{eq:sigmoid}, as well as the \textit{hyperbolic tangent} function defined in Eq.~\eqref{eq:hyp_tang}:
\begin{align}
	\sigma(x) &= \frac{1}{1 + \exp(-x)} \label{eq:sigmoid} \\
	\text{tanh}(x) &= \frac{ \exp(x) - \exp(-x) }{\exp(x) + \exp(-x)} \label{eq:hyp_tang}.
\end{align}
A commonly utilized unbounded activation function is the \textit{rectified linear unit}, commonly called \textit{ReLU}, defined in Eq.~\eqref{eq:relu}:
\begin{align}
	\text{ReLU}(x) &= \max(0, x). \label{eq:relu}
\end{align}
An artificial neural network consists of layers of ``stacked'' artificial neurons, where the output of each layer of neurons acts as the inputs to the next layer of neurons, as depicted in Fig.~\ref{fig:neural_net}. The input layer is simply represented as a set of constant values and is correspondingly represented as a lighter shade in Fig.~\ref{fig:neural_net}, while the neurons in the output layer still maintain the complete neural structure of weights, biases, and an activation function as in Fig.~\ref{fig:neuron}. Notice that the number of neurons in the output layer is equivalent to the dimension of the desired output from the neural network.

The layers of neurons which do not correspond to the input layer or the output layer are called the \textit{hidden layers} of the network. The number of neurons in each hidden layer can be selected by the designer of the network architecture. Additionally, the neural network is typically \textit{fully connected}, where each neuron from the previous layer is input to all neurons in the following layer. The parameters of the neural network, i.e. the weights and biases of each artificial neuron in the network, are optimized, or ``learned,'' during training via any desired optimization algorithm~{\citep{sun2019survey,kochenderfer2019algorithms}}. Several optimization algorithms which are commonly utilized in the field of machine learning include stochastic gradient descent~\citep{robbins1951stochastic}, Adagrad~{\citep{duchi2011adaptive}}, RMSProp~{\citep{hintonlecture}}, and Adam~\citep{kingma2014adam}.

For more background on neural network architectures or general machine learning concepts, the authors recommend referencing \citep{bishop2006pattern}~and~{\citep{bengio2017deep}} for a valuable and accessible introduction.

%% file: sections/acknowledgements.tex
\section*{Acknowledgements}\label{sec:ack}
This material is based upon work supported by the Air Force Research Lab (AFRL) at Kirtland Air Force Base in Albuquerque, New Mexico under grant number FA9453-20-1-0002. This material is additionally based upon work supported by the National Science Foundation Graduate Research Fellowship under grant number DGE-1656518. Any opinions, findings, and conclusions or recommendations expressed in this material are those of the author(s) and do not necessarily reflect the views of the Air Force Research Lab or the National Science Foundation.